\documentclass[a4paper,10pt]{article}
\usepackage[utf8]{inputenc}

\usepackage{amssymb}
\usepackage{amsmath}
\usepackage{amsthm}
\usepackage{mathtools}
\usepackage{natbib}
\usepackage{subfig}
\usepackage{graphicx}
\usepackage{extarrows}
\usepackage{multirow}
\usepackage{hyperref}  
\usepackage{float}
\usepackage{caption} 
\usepackage{array}
\usepackage{framed}
\usepackage{dsfont}
\usepackage{xcolor}
\usepackage{slashbox}
\usepackage{booktabs}
\usepackage{siunitx}

\newtheorem{definition}{Definition}[subsection]\newtheorem{theorem}[definition]{Theorem}

\title{{\bf A Comparison of Single and Multiple Changepoint Techniques for Time Series Data}}

\author{Xueheng Shi, Colin Gallagher, Robert Lund and Rebecca Killick}

\begin{document}

\maketitle

\begin{abstract}
This paper describes and compares several prominent single and multiple changepoint techniques for time series data. Due to their importance in inferential matters, changepoint research on correlated data has accelerated recently.  Unfortunately, small perturbations in model assumptions can drastically alter changepoint conclusions; for example, heavy positive correlation in a time series can be misattributed to a mean shift should correlation be ignored.  This paper considers both single and multiple changepoint techniques.  The paper begins by examining cumulative sum (CUSUM) and likelihood ratio tests and their variants for the single changepoint problem; here, various statistics, boundary cropping scenarios, and scaling methods (e.g., scaling to an extreme value or Brownian Bridge limit) are compared. A recently developed test based on summing squared CUSUM statistics over all times is shown to have realistic Type I errors and superior detection power.  The paper then turns to the multiple changepoint setting.  Here, penalized likelihoods drive the discourse, with AIC, BIC, mBIC, and MDL penalties being considered. Binary and wild binary segmentation techniques are also compared. We introduce a new distance metric specifically designed to compare two multiple changepoint segmentations.  Algorithmic and computational concerns are discussed and simulations are provided to support all conclusions.  In the end, the multiple changepoint setting admits no clear methodological winner, performance depending on the particular scenario.  Nonetheless, some practical guidance will emerge. 
\end{abstract}

\section{Introduction}
Changepoints (abrupt shifts) arise in many time series due to changes in recording equipment, observers, etc.  In climatology, temperature trends computed from raw data can be misleading if homogeneity adjustments for station relocation moves and gauge changes are not {\em a priori} made to the record. \citet{LuLund2007} give an example where trend conclusions reverse when changepoint information is neglected.  Cases with multiple changepoints are also frequently encountered; for example, in climatology, United States weather stations average about six station moves and/or gauge changes per century of operation \citep{Menne_et_al2009}.

This paper intends to guide the researcher on the best changepoint techniques to use in common time series scenarios. Assumptions are crucial in changepoint analyses and can significantly alter conclusions; here, correlation issues take center stage.  It is known that changepoint inferences made from positively correlated series can be spurious if correlation is not taken into account.  Even lag one correlations as small as 0.25 can have deleterious consequences on changepoint conclusions \citep{Lund_etal_2007}. 

This paper's primary contribution is to extend/modify many of the popular changepoint methods for IID data to correlated settings. Much of our work lies with developing methods that put all techniques, to the best extent possible, on the same footing in time series settings.  For example, we will see that single changepoint tests generally work best when applied to estimated versions of the series' one-step-ahead prediction residuals, computed under a null hypothesis of no changepoints. Because of this, tests that handle one-step-ahead prediction residuals need to be developed.  Two other novel contributions in this article are: (1) developing and proposing a new single changepoint test based on the square of the cummulative sum of one-step-ahead prediction residuals (see Section \ref{NEWTEST}), and 2) presenting a new distance that compares multiple changepoint segmentations (see Section \ref{metric}).  The comparative aspect of the paper is yet another contribution --- and there is much to compare.  In addition to comparing different statistics via Type I errors and powers, the paper also compares different asymptotic scaling methods.

Academic changepoint research commenced with the single changepoint case for independent and identically distributed (IID) data in \citet{page1955}.  The subject is now vast, with hundreds of papers devoted to the topic.  With our lofty objectives, some concessions are necessary.  Foremost, this paper examines mean shift changepoints only; that is, while series mean levels are allowed to abruptly shift, the variances and correlations of the series are held constant (stationary) in time.  Changepoints can also occur in variances (volatilities) \citep{CEK2020}, in the series' correlation structures \citep{Davis_etal_2006}, or even in the marginal distribution of the series \citep{GLR_2012}.  Secondarily, the simulation results reported here are for Gaussian series only.  Thirdly, we compare the most common types of techniques within the literature, notably discluding those based on energy statistics \citep{MattesonJames2014}, moving sums \citep{mosum}, and U statistics \citep{Dehling2015}.

The rest of this paper proceeds as follows.  Section \ref{sec:AMOC} overviews single changepoint detection methods, typically referred to as at most one changepoint (AMOC) tests.  Here, a variety of test statistics and their scalings are reviewed and adapted to the time series setting. We specifically discuss two methods for modifying changepoint techniques based on IID data: 1) retain the IID test statistic and modify the limiting distribution for any correlation; and 2) modify the test statistic to account for the correlation.  Section \ref{sec:AMOCsims} compares AMOC detectors in a simulation study. Thereafter, we move to the case of multiple changepoints, where performance assessment becomes more challenging.  Here, a novel changepoint configuration distance specifically designed for our comparisons is developed.  Simulations in Section \ref{sec:multiple} consider a variety of multiple changepoint configurations.  We summarize results in Section \ref{sec:conc} with recommendations for practitioners.

\section{Single Changepoint Techniques}
\label{sec:AMOC}
Let $\{X_t\}_{t=1}^N$ be the observed time series and $\gamma(h) = \mbox{Cov}(X_{t+h}, X_t)$ be the lag $h$ autocovariance of the series.  We wish to test whether there exists a change in the mean structure {\itshape while assuming the second order structure is constant over time}.  An AMOC model with the changepoint occurring at the unknown time $k$ is
\begin{align}
X_t=
    \begin{cases}
    \mu + \epsilon_t, \qquad &\text{for $1 \le t \le k$,}\\
    \mu + \Delta +\epsilon_t, \qquad &\text{for $k+1 \le t \le N$,}
    \end{cases}\label{eqn:model}
\end{align}
where $\mu$ is an unknown location parameter, $\Delta$ is the magnitude of mean shift at time $k$, and $\{\epsilon_t \}$ is a stationary time series with zero mean and lag $h$ autocovariance $\gamma(h)$.  A hypothesis test for this scenario is:
\begin{equation}
    H_0: \Delta=0 \quad \text{versus} \quad H_1: \Delta  \neq 0 \quad \text{for some}~k \in \{ 1, \ldots, N-1 \}.
\end{equation}\label{eqn:scpt_hypo}

When the $\{ \epsilon_t \}$ are independent, cumulative sum (CUSUM) and likelihood ratio tests (LRT) are well understood, see \cite{csorgo1997limit} and \cite{chengupta2011}.  When incorporating general stationary autocovariance aspects into a changepoint testing framework, there are two common strategies: 1) keep the IID test statistic and identify any changes in the limiting distribution induced by the correlation; and 2) incorporate the autocovariance within the test statistic. \cite{ANTOCH1997} provide a summary of the first approach for many common changepoint statistics and provide simulations indicating how autocorrelation impacts the performance of the hypothesis tests; \cite{Kirch2007} uses resampling techiques to improve the finite sample performance of these tests.  \cite{robbins2011} shows that estimating and using the autocorrelation (the second approach) is preferable with CUSUM and LRTs.

\subsection{CUSUM Tests}
The CUSUM method was first introduced by \cite{page1955} and compares sample means before and after each admissible changepoint time via the statistic
\begin{align}
        \max_{1 \leq k < N} \left|\text{CUSUM}_X(k) \right| :=
\max_{1 \leq k < N}\left|\frac{1}{\sqrt{N}} \left[ \sum_{t=1}^{k} X_t - \frac{k}{N} \sum_{t=1}^N X_t \right]\right|. 
\label{eqn:maxcusumx}
\end{align}
CUSUM tests have relatively poor detection power when the changepoint occurs near the boundaries (times 1 or $N$).  Conversely, false detection is more likely to be signaled near the boundaries (i.e., when one of the segment sample means has a comparatively high variance). Because of this, cropped-CUSUM methods, which weight or ignore observations close to the two boundaries, were developed. Simulations for cropped settings analogous to those below are presented in the supplementary material; in general, one loses power by cropping. See \cite{csorgo1997limit} for generalities on cropping.

In our first scenario, where the IID test statistic described in \eqref{eqn:maxcusumx} is used, its asymptotic distribution for correlated data, under the null hypothesis of no changepoints, is known from \cite{Macneill1974} and \cite{csorgo1997limit}. 
\begin{theorem}
\label{thm:CUSUM}
\citep{csorgo1997limit}

Assume that $\{X_t\}$ follows \eqref{eqn:model}, $\{\epsilon_t\}$ has the usual causal linear representation, $\epsilon_t=\sum_{i=0}^{\infty}\psi_i Z_{t-i}$, and $\hat{\eta}^2$ is a null hypothesis based consistent estimator of $\eta^2$, the long-run variance parameter
\begin{equation}
\label{eq:lrcov}
    \eta^2:=\lim_{n \rightarrow \infty} \frac{1}{n} \mbox{Var}\left(\sum_{t=1}^n \epsilon_t \right).
\end{equation}
Then under $H_0$,
\begin{equation}\label{eq:lrteststat}
        \frac{1}{\hat{\eta}} \max_{1 \leq k < N} \left| \text{CUSUM}_X(k) \right|\; \xrightarrow{\; {\cal D} \;} \; 
        \sup_{t \in [0,1] } |B(t)|.
\end{equation}
Here, it is assumed that $\{Z_t\}$ is IID with zero mean, variance $\sigma^2$, a finite fourth moment, and $\sum_{j=0}^{\infty} |\psi_j|< \infty$.   Moreover, $\{B(t), t \in [0,1] \}$ denotes a standard Brownian bridge process obeying $B(t) = W(t) - tW(1)$, where $\{ W(t), t \ge 0 \}$ is a standard Wiener process.
\end{theorem}

Theorem \ref{thm:CUSUM} requires estimation of $\eta^2$, which is challenging by itself \citep{stoica2005}.

While this result provides an appropriate asymptotic test, strong correlation often degrades CUSUM performance \citep{robbins2011}. That is, convergence to the limit law is faster for independent data than for positively correlated data.  As such, it is often beneficial to decorrelate heavily dependent data before using CUSUM methods.  This brings us to our second approach, which incorporates the correlation within the test statistic. For CUSUM methods, this is achieved by replacing the data by one-step-ahead linear prediction residuals.

The autoregressive moving average (ARMA) one-step-ahead linear prediction residuals are defined as:
\begin{equation}
    \hat{Z}_t = \Dot{X}_t-\hat{\phi}_1\Dot{X}_{t-1}-\cdots -\hat{\phi}_p\Dot{X}_{t-p}-
    \hat{\theta}_1\hat{Z}_{t-1}-\cdots-\hat{\theta}_q\hat{Z}_{t-q},
\label{eq:ARMAresid}
\end{equation}
where $\Dot{X}_t=X_t-\bar{X}$, $\bar{X}= N^{-1}\sum_{t=1}^N X_t$, and $\phi_1, \ldots , \phi_p$ are the autoregressive coefficients and $\theta_1, \ldots, \theta_q$ are the moving-average coefficients.  Here, the edge conditions take $\Dot{X}_t=\hat{Z}_t=0$ for any $t<0$. The estimator $\hat{\sigma}^2 = N^{-1} \sum_{t=1}^N \hat{Z}_t^2$ is used to estimate the variance of $Z_t$. The residual CUSUM statistic at time $k$ is
\begin{equation}
    \max_{1 \leq k < N}
    \left|\text{CUSUM}_Z(k)\right| := \max_{1 \leq k < N}\left|\frac{1}{\sqrt{N}}\left(\sum_{t=1}^{k}\hat{Z}_t - \frac{k}{N}\sum_{t=1}^N\hat{Z}_t\right)\right|,
\label{eq:maxCUSUMresid}
\end{equation}
where our notational convention appends the subscript $Z$ to indicate use of prediction residuals.

The asymptotic distribution of the CUSUM of the one-step-ahead prediction residuals was studied in \cite{robbins2011}.

\begin{theorem}
\citep{robbins2011}
\label{thm:cusumz}

Suppose that $\{\epsilon_t\}$ is a causal and invertible ARMA series with IID $\{ Z_t \}$ having zero mean, variance $\sigma^2$, and with $E[Z_t^4] < \infty$.  Let $\{ \hat{Z}_t \}$ be the estimated one-step-ahead prediction residuals in (\ref{eq:ARMAresid}).  Then under the null hypothesis of no changepoints, 
\begin{equation}
    \frac{1}{\hat{\sigma}} \max_{1 \leq k < N} |\mbox{CUSUM}_Z(k)| -
    \frac{1}{\hat{\eta}}   \max_{1 \leq k < N} |\mbox{CUSUM}_X(k)| = o_p(1),
\end{equation}
when all ARMA parameters and $\eta^2$ are estimated in a $\sqrt{N}$-consistent manner. It hence follows that
\begin{align}
\label{cusumz}
    \frac{1}{\hat{\sigma}} \max_{1 \leq k < N} |\mbox{CUSUM}_Z(k)|\; 
    \xrightarrow{\;  {\cal D} \;} \; \sup_{0 \le t \le 1} |B(t)|.
\end{align}
\end{theorem}

Both of these approaches are compared in Section \ref{sec:AMOCsims}.

\subsection{SCUSUM Tests}
\label{NEWTEST}
As an alternative to using partial sums to detect mean shifts, several authors have considered summing the squares of these partial sums.  The resulting test statistic converges to the integral of the square of a Brownian Bridge. With SCUSUM denoting the test's acronym, for IID data, the test statistic is
\begin{align}
 \label{eqn:scusum}
 \text{SCUSUM}_X := \frac{1}{N}\sum_{k=1}^N \left[ 
 \frac{\text{CUSUM}_X(k/N)}{\hat{\sigma}} \right]^2.
\end{align}
The squared CUSUM (SCUSUM) test does not by itself yield an estimate of the changepoint location.  If the SCUSUM test indicates that a changepoint is preferred, then its location is estimated as that argument(s) that maximizes the absolute CUSUM statistic.

We again consider two approaches for modifying the SCUSUM test for correlation. First, the distribution of the statistic in \eqref{eqn:scusum} for autocorrelated data under the null hypothesis can be quantified.  The following result follows from Theorem \ref{thm:CUSUM} via an application of the continuous mapping theorem.

\begin{theorem}

Assume that $\{X_t\}$ follows \eqref{eqn:model}, $\{\epsilon_t\}$ has the causal linear representation assumed in Theorem 1, and $\hat{\eta}^2$ is a null hypothesis based consistent estimator of $\eta^2$, the long-run variance defined in \eqref{eq:lrcov}. Then under $H_0$,
\begin{align}
    \text{SCUSUM}_X=\frac{1}{N}\sum_{k=1}^N \left[ \frac{\text{CUSUM}_X(k/N)}{\hat{\eta}} \right]^2 \xlongrightarrow{{\cal D}} \int_0^1 B^2(t) dt.
\end{align}
\end{theorem}

Our second approach for incorporating correlation uses the one-step-ahead prediction residuals in place of the original data.  The SCUSUM test statistic for this scheme is
\begin{align}
 \label{eqn:scusumz}
 \text{SCUSUM}_Z := \frac{1}{N}
 \sum_{k=1}^N 
 \left[ \frac{\text{CUSUM}_Z(k/N)}{\hat{\sigma}} 
 \right]^2.
\end{align}

The asymptotic distribution of \eqref{eqn:scusumz} can be derived from Theorem \ref{thm:cusumz} via the continuous mapping theorem.

\begin{theorem}

With $CUSUM_Z$ defined as in Theorem \ref{thm:cusumz}, under the null hypothesis of no changepoints, 
\begin{align}
    \text{SCUSUM}_Z=\frac{1}{N}\sum_{k=1}^N \left[ \frac{\text{CUSUM}_Z(k/N)}{\hat{\sigma}} \right]^2 \xlongrightarrow{ {\cal D}} \int_0^1 B^2(t)\;dt.
\end{align}
\end{theorem}

The distribution of $\int_0^1 B(t)^2 dt$ was investigated in \cite{tolmatz2002}. We note that \cite{bai1993partial} proposed using the sum of the square of partial sums of ARMA residuals to detect a single changepoint in autocorrelated data; this test statistic converges to the integral of a squared {\em Brownian Motion} rather than the integral of the square of a {\em Brownian Bridge}. To our knowledge, the variant in (\ref{eqn:scusumz}) has not previously been proposed nor studied in the literature.

The differences between CUSUM and CUSUMz statistics were investigated in \cite{robbins2011} and their simulations indicate that the latter statistic is superior to the former in terms of type I error and power.  Our simulations confirm this finding.  As such, in the remainder of the paper, we do not consider SCUSUM (without the subscript $Z$) tests further.

\subsection{Likelihood Ratio Tests}
While CUSUM tests are non-parametric, LRTs are inherently parametric.  Several error distributions have been considered by previous authors, by far the most common being normal --- this is the distribution considered here.

The LRT compares the likelihood under the null hypothesis to likelihoods under alternatives with a changepoint.  The LRT statistic for a changepoint has the general form
\begin{equation}
\Lambda=\max_{1 \leq k < N} \Lambda_k, \quad
\quad
\Lambda_k=
\frac{L_0(\hat{\mu}_0)}
{L_k(\hat{\mu}_1, \hat{\mu}_2)},
\end{equation}
where $L_0$ denotes a null hypothesis likelihood and $L_k$ an alternative likelihood when the changepoint occurs at time $k$.  Elaborating, $\hat{\mu}_0$ is the maximum likelihood estimator (MLE) for $E[ X_t ]$ under $H_0$, and $\hat{\mu}_1$ and $\hat{\mu}_2$ are the MLEs for the means of the two segments under the alternative when there is a mean shift at time $k$. The end statistic is then the maximum over all admissible changepoint locations $k$.   When correlation exists in $\{ X_t \}$, the form of the Gaussian likelihood can be found in \cite{Brockwell_Davis_91}; this form may contain additional ARMA or other correlation parameters that have to be estimated. 

When the errors are from a causal and invertible Gaussian ARMA process, \cite{jandhyala2013} develop asymptotics, scaling to an extreme value limit.  While the asymptotics require one to estimate the ARMA parameters in calculation of the $\Lambda_k$ statistics, the limit distribution does not depend on the ARMA parameters, nor does the scheme require any cropping of the boundary times. 

\begin{theorem}
\citep{jandhyala2013}

Suppose that $\{\epsilon_t\}$ is a causal and invertible ARMA series with IID $\{ Z_t \}$ satisfying the assumptions in Theorem 2.  Then the LRT statistic is
\begin{equation}
\label{LRT}
U = \max_{1 \leq k < N} 
\left( -2 \log(\Lambda_k) \right), \quad
\Lambda_k = \left(\frac{\hat{\sigma}_k^2}{\hat{\sigma}_{H_0}^2} \right)^{\frac{N}{2}}. 
\end{equation}
Here, $\hat{\sigma}^2_k$ is the MLE estimate of the ARMA white noise process variance when there is a changepoint at time $k$ and $\hat{\sigma}^2_{H_0}$ is an estimate of this same variance under the null hypothesis of no changepoints.  This statistic can be scaled to a Gumbel extreme value limit:
\[
W_U := \sqrt {2 U \log \log(N)}- \left[2\log \log(N) + \frac{1}{2} \log \log \log(N) - \frac{1}{2} \log \pi \right].
\]
Then under $H_0$,
\begin{equation}
\label{eq:Gumbel}
\lim_{N \rightarrow \infty} 
\mathbb{P}(W_U \le x) = \exp(-2\exp(-x) ), \qquad -\infty < x < \infty.
\end{equation} 
Specifically, $H_0$ is rejected when $W_U$ is too large to be explained by the distribution in (\ref{eq:Gumbel}).
\end{theorem}

Another way of scaling the $\Lambda_k$ statistics involves cropping boundary times.   Like the CUSUM test, the LRT is volatile at times near the boundaries. In fact, $\Lambda \stackrel{{\cal D}} {\rightarrow} \infty$ as $N \rightarrow \infty$  should the maximum be taken over the entire range $1 \leq k < N$ under the null hypothesis of no changepoints.  A common cropped LRT simply truncates admissible times near the boundaries; for example, with $0 < \ell < h < 1$, $\ell$ being close to zero and $h$ being close to unity, set
\begin{equation}
    U_{{\rm crop}}=\max_{\ell \le k/N \le h} (-2\log(\Lambda_k)).
\end{equation}
\cite{robbins2011} shows that
\begin{equation}
\label{LRTc}
    U_{{\rm crop}} \xlongrightarrow{{\cal D}} \sup_{\ell \le t \le h} \frac{B^2(t)}{t(1-t)}.
\end{equation}

As the next section shows, LRTs are not competitive in changepoint detection problems. While simulations are presented for the above extreme value test in the next section, simulations for cropped LRTs are delegated to the supplementary material --- both methods perform poorly.  

As a final comment here, deriving a LRT test for independent data, and then replacing the data with one-step-ahead prediction residuals, another avenue for dealing with dependence, does not yield a methodologically distinct path.  Specifically, if one derives a LRT statistic for independent series and then substitutes one-step-ahead prediction residuals in place of the original data, the limit law in (\ref{LRTc}) again arises.  The boundaries again must be cropped to ensure a proper limiting distribution.  The discussion around (1.4.22) --- (1.4.27) in \citep{csorgo1997limit} provides more detail on this route; 
see also \cite{LavielleMoulines2000} for more on LRTs for correlated data. 

\section{AMOC Simulations}
\label{sec:AMOCsims}

This section investigates the finite sample performance of the Section \ref{sec:AMOC} tests (cropped CUSUMz, CUSUMz, SCUSUMz, LRT) through simulation. Results for the cropped test statistics are delegated to the supplementary material; results for the others are presented here.

Desirable tests have reasonable (non-inflated) false detection rates when no changepoints exist, and large detection powers when a changepoint is present, regardless of the degree of correlation.  For each statistic under consideration, the impact of autocorrelation on the Type I error is first explored. We then examine detection powers of the tests when a changepoint exists.  First order Gaussian autoregressions (AR(1)) are considered here with $\sigma^2=1$; other structures are examined in the supplementary material.  

Figure \ref{fig:t1error} summarizes results for $N=1,000$ across varying AR(1) correlation parameters $\phi$.  Our conclusions do not vary for different $N$ --- see the supplementary material.  Figure \ref{fig:t1error} shows that the only method to retain a controlled Type I error across all $\phi$ is the SCUSUMz.  The LRT is the worst performing method, being far too conservative for all $\phi$, except $0.95$, when it becomes highly inflated.  The CUSUMz method is also slightly conservative, becoming more so as $\phi$ increases.  Since we are using the asymptotic distribution for each of the 10,000 test statistics, we would expect the $0.05$ type I error to be reasonably maintained.

\begin{figure}[H]
\centering
\includegraphics[scale=0.5]{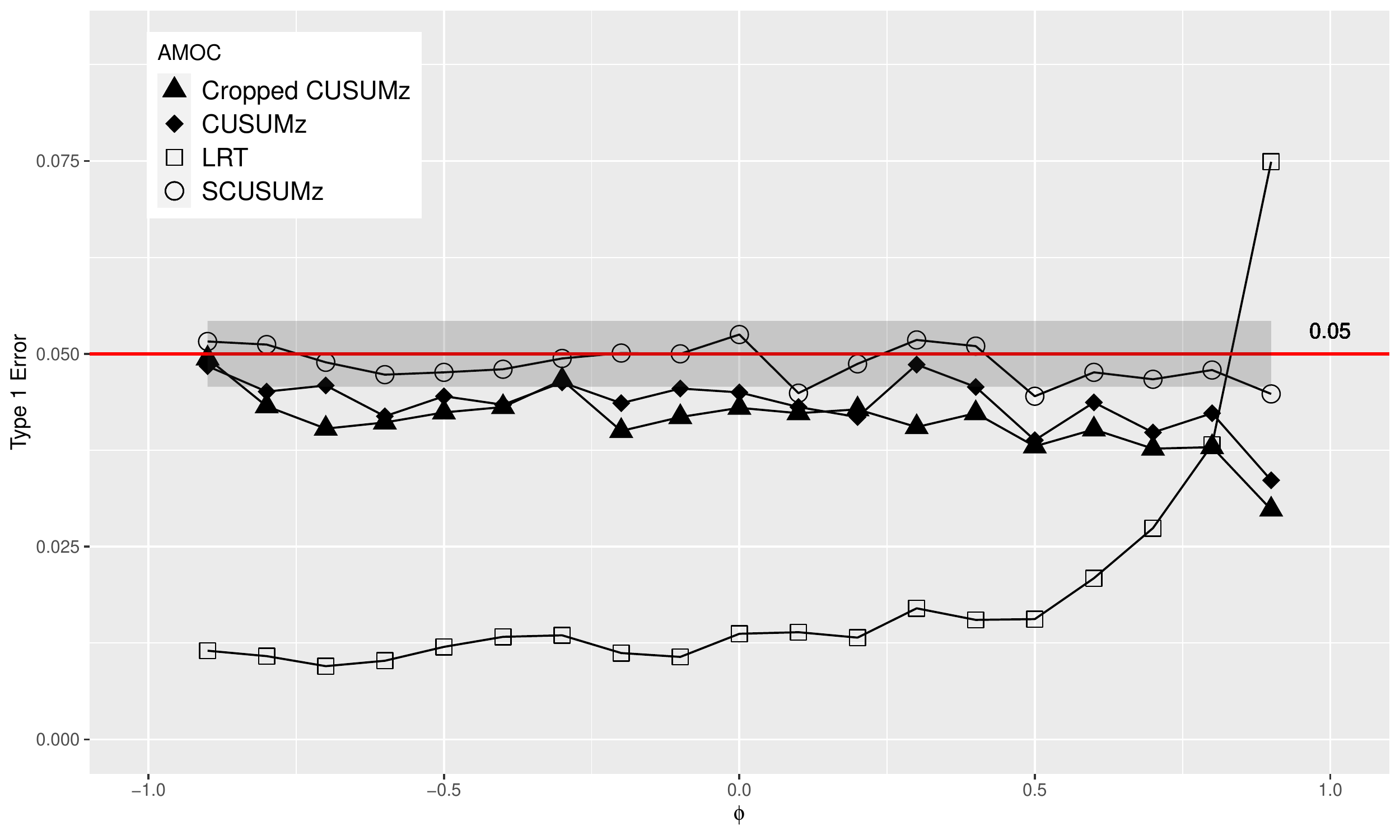}
\caption{Type I Errors for an AR$(1)$ Series with Different $\phi$ When $N=1000$.}
\label{fig:t1error}
\end{figure}

\begin{figure}[H]
\centering
\includegraphics[scale=0.5]{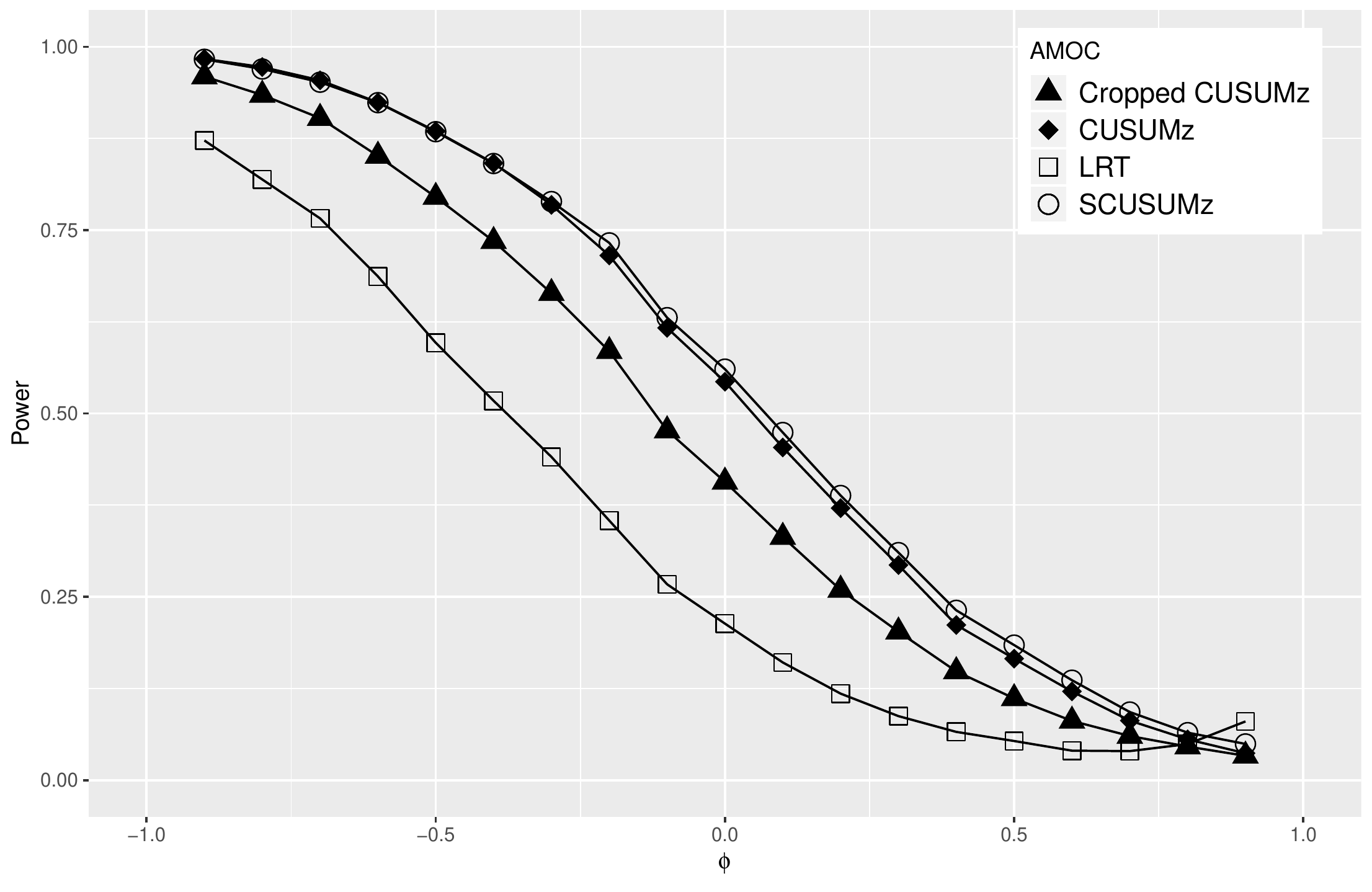}
\caption{Detection Powers for an AR$(1)$ Series with Different $\phi$.  Here, $N=1,000$ and $\Delta=0.15$.}
\label{fig:power_Delta015}
\end{figure}

\begin{figure}[H]
\centering
\includegraphics[scale=0.5]{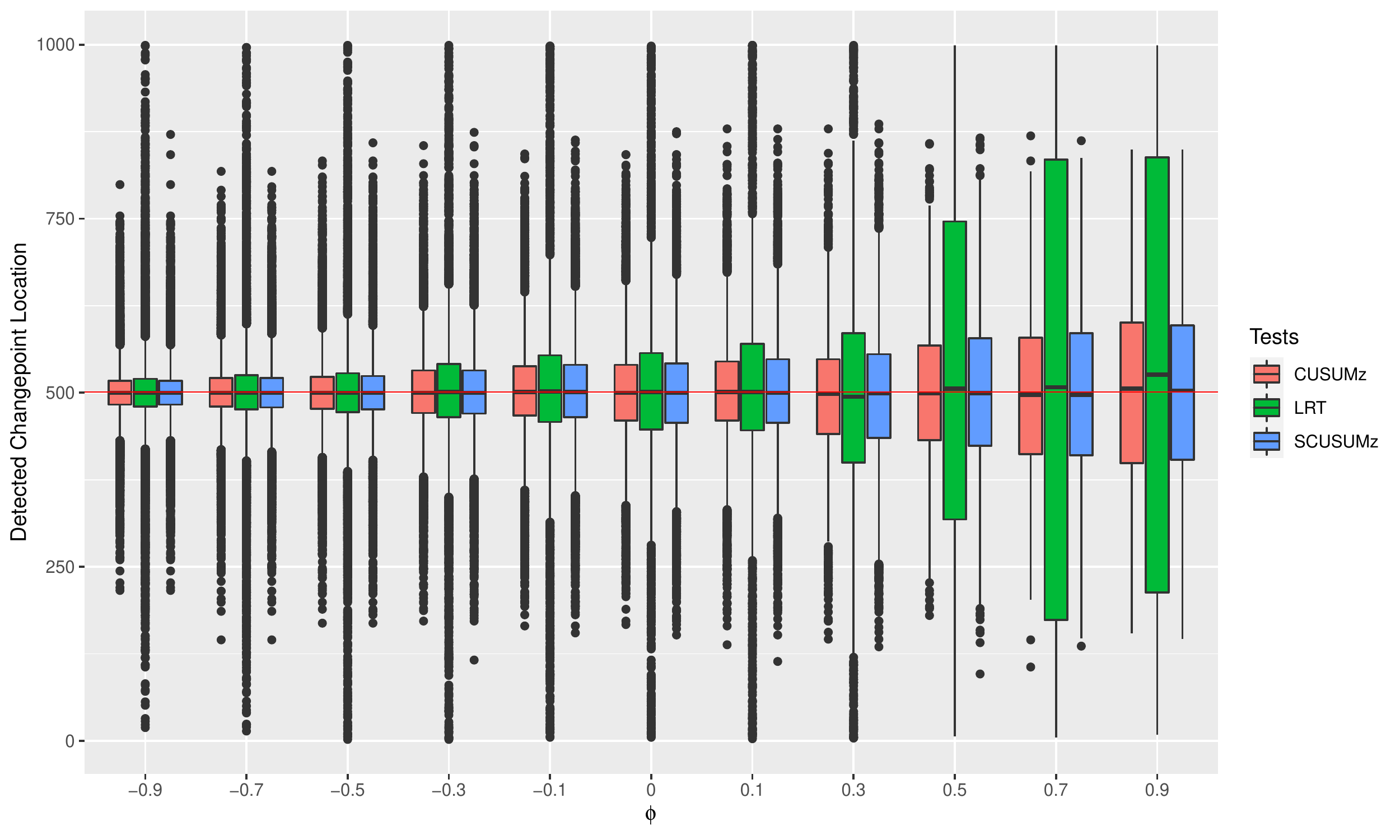}
\caption{Boxplots of Detected Changepoint Locations for an AR$(1)$ Series with Different $\phi$.  Here, $N=1,000$ and $\Delta=0.15$.}
\label{fig:scpt_location_delta015}
\end{figure}

\begin{figure}[H]
\centering
\includegraphics[scale=0.5]{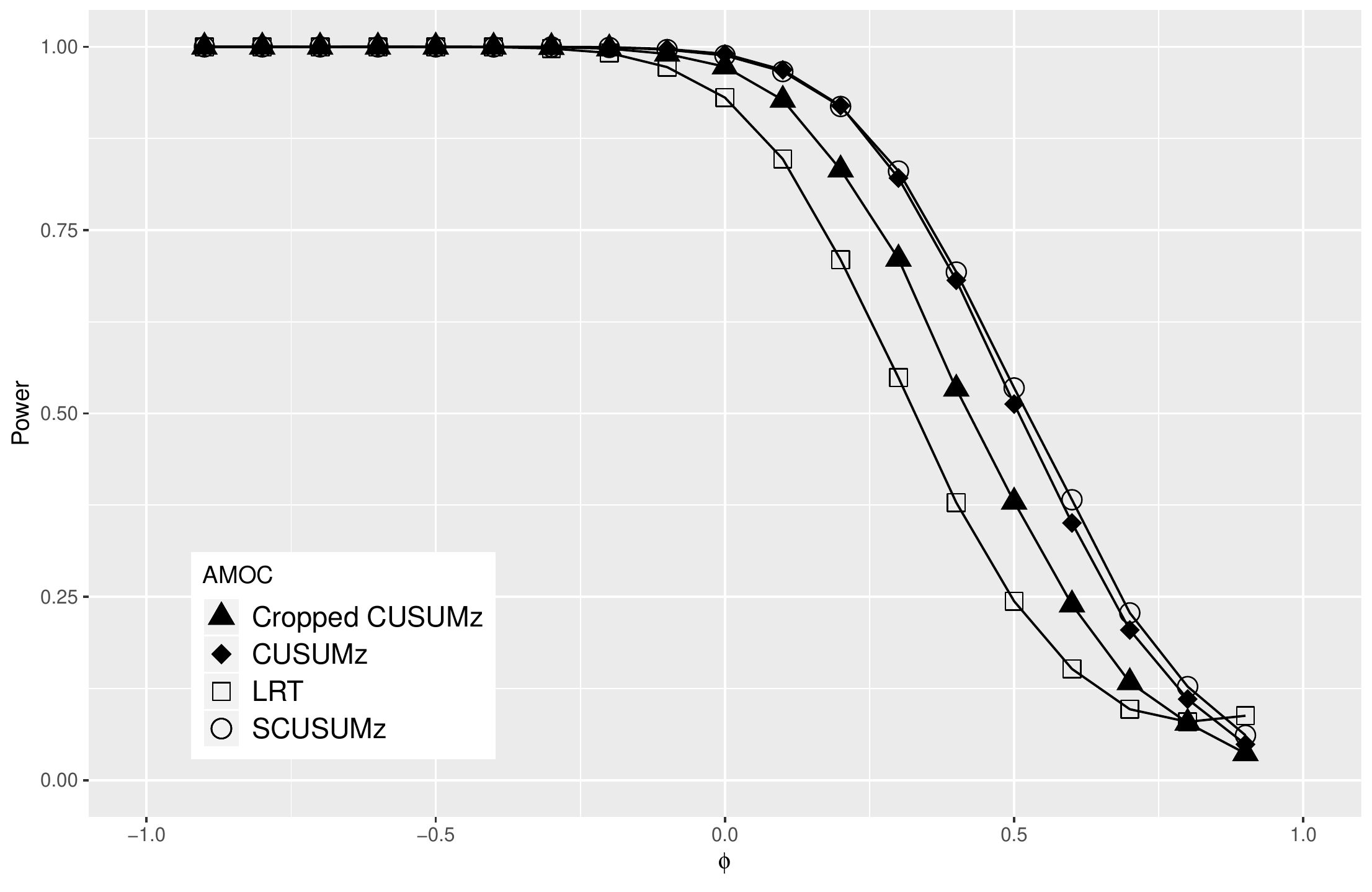}
\caption{Detection Power for an AR$(1)$ Series with Different $\phi$.   Here, $N=1000$ and $\Delta=0.3$.}
\label{fig:power_Delta03}
\end{figure}

\begin{figure}[H]
\centering
\includegraphics[scale=0.5]{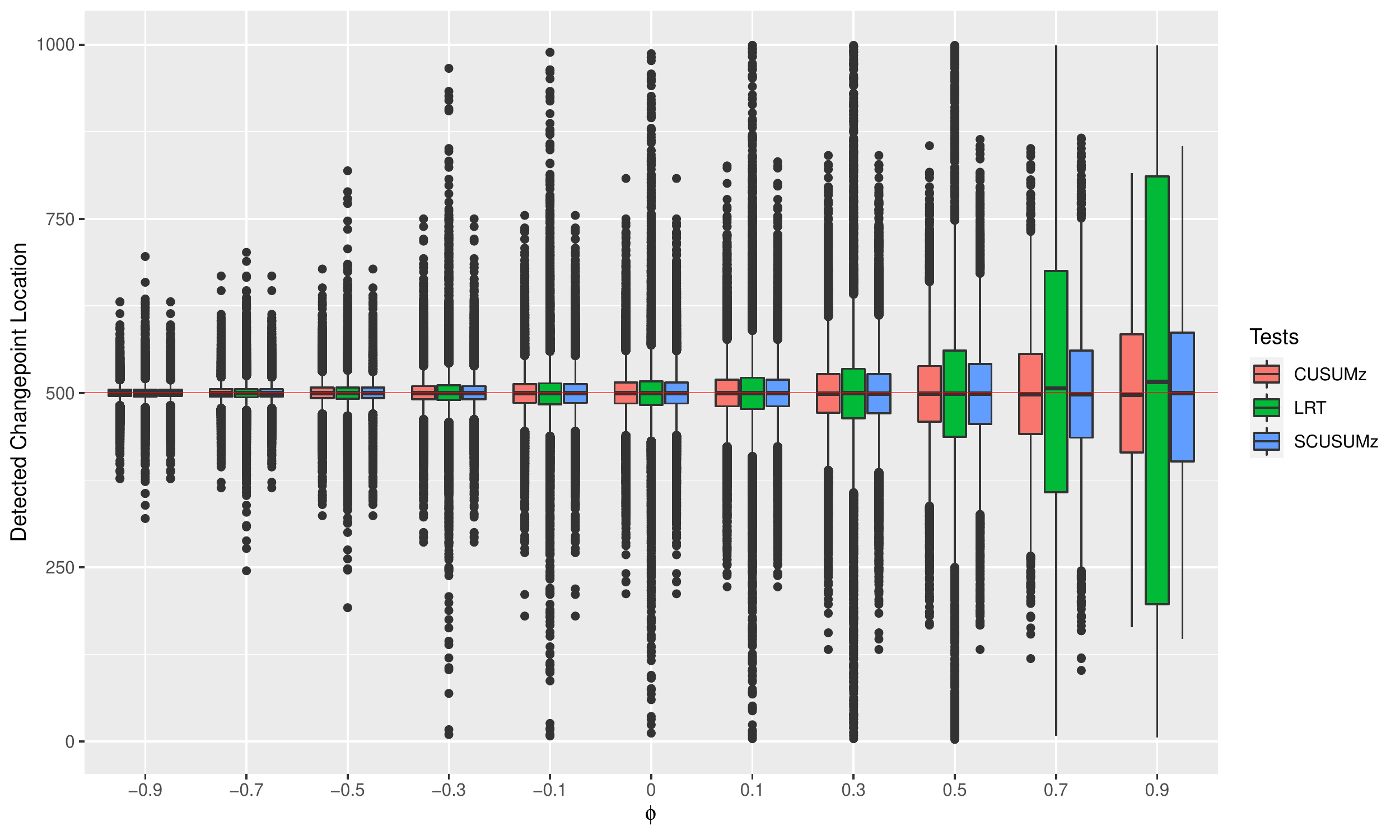}
\caption{Detected Changepoint Location for an AR$(1)$ Series with Different $\phi$.  Here, $N=1000$ and $\Delta=0.3$.}
\label{fig:scpt_location_delta03}
\end{figure}

We now consider test detection powers.  In general, the detection power of an AMOC test depends on the degree of correlation, the size of the mean shift, and the location of the changepoint time \citep{robbins-JASA}. Figures \ref{fig:power_Delta015} ($\Delta=0.15$) and \ref{fig:power_Delta03} ($\Delta=0.3$) show empirical powers based on $10,000$ independent Gaussian simulated series of length $N=1,000$.  Sample powers are plotted as a function of $\phi$ when the mean shift lies in the center of the series (time $501$).  The figures demonstrate the drastic effects of autocorrelation on the power of changepoint tests.  While the LRT had the highest empirical power when $\phi=.95$, the estimated changepoint location of LRT is biased and more variable than that for the $\text{CUSUM}_Z$ and $\text{SCUSUM}_Z$ tests, see Figures \ref{fig:scpt_location_delta015} and \ref{fig:scpt_location_delta03}. The LRT test also has a Type I error far exceeding 0.05; as such, it's higher power does not imply better overall performance. Overall, the $\text{CUSUM}_Z$ and $\text{SCUSUM}_Z$ tests are more powerful than the others.  Note also that $\text{SCUSUM}_Z$ has higher power than $\text{CUSUM}_Z$ for each $\phi$ considered.  Additional simulations (not shown) duplicate this conclusion for other sample sizes.  The SCUSUMz statistic is clearly the best test. 
 
\begin{figure}[htb]
\centering
\includegraphics[scale=0.5]{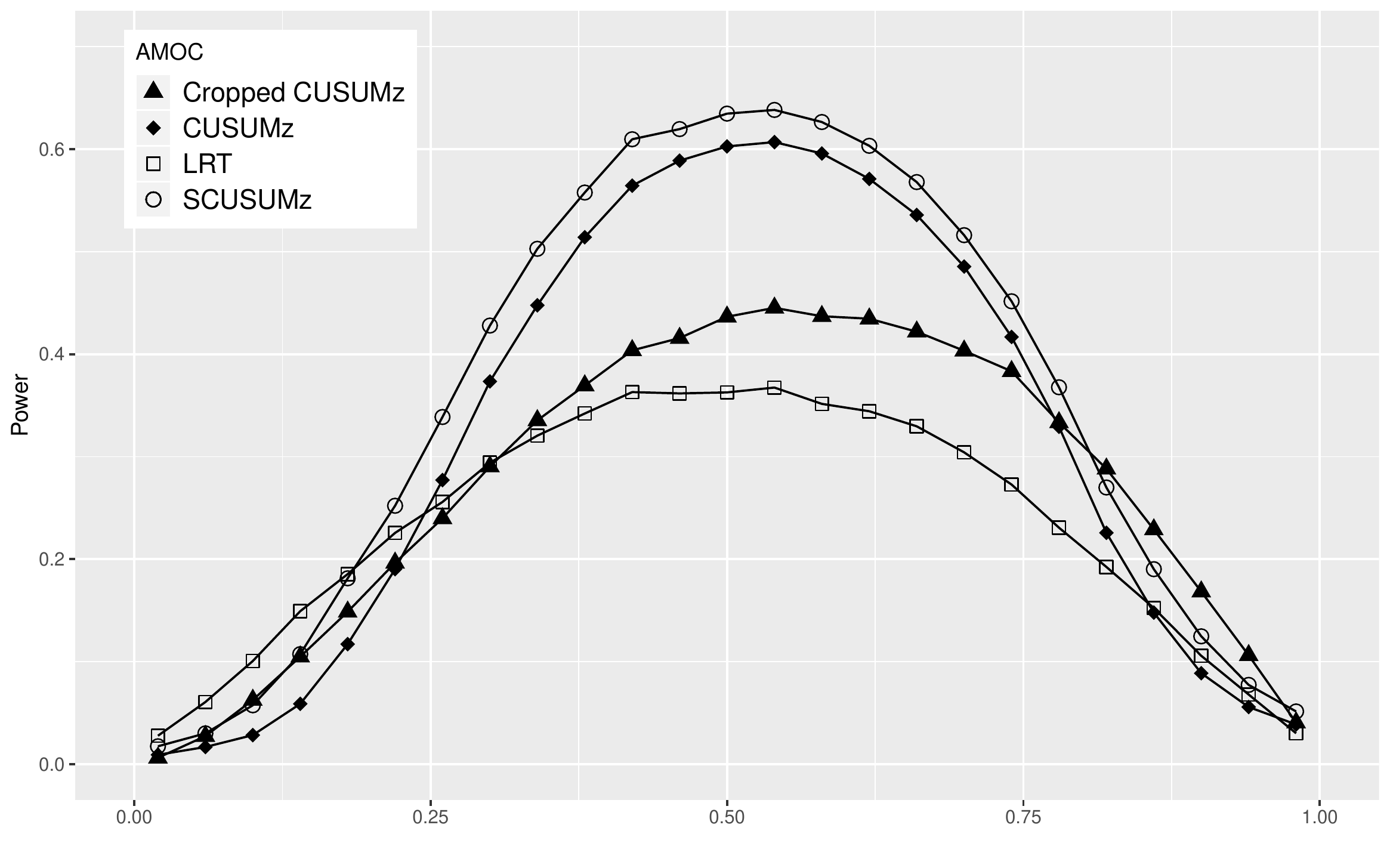}
\caption{A Graph of $\frac{\tau}{N}$ Against Power with $N=500$ and $\Delta=0.5$ for an AR$(1)$ Series with $\phi=0.5$.}
\label{fig:power_by_tau}
\end{figure}

Finally, we examine the effect of the changepoint location. Simulation specifications are as in the above paragraph, but the location of the changepoint is now varied and $\phi$ is fixed as 0.5.  Figure \ref{fig:power_by_tau} displays empirical powers.   The largest detection powers occur when the changepoint is near the center of the record, as expected, with power decreasing as the changepoint time moves towards a boundary. The $\text{SCUSUM}_Z$ appears to be the most accurate overall. However, the LRT test is preferable when the changepoint occurs near the beginning of the record.

\section{Multiple Changepoint Techniques}
\label{sec:multiple}

Now suppose that $\{X_t\}_{t=1}^N$ has an unknown number of changepoints, denoted by $m$, occurring at the unknown ordered times $1 < \tau_1 < \tau_2 < \cdots < \tau_m \le N$. Boundary conditions take $\tau_0=0$ and $\tau_{m+1}=N$. These $m$ changepoints partition the series into $m+1$ distinct regimes, the $i^{th}$ regime having its own distinct mean and containing the data points $\{ X_{\tau_i+1}, \ldots, X_{\tau_{i+1}}\}$. The model can be written as $X_t=\kappa_{t} + \epsilon_t$, where $\kappa_t = \mu_{r(t)}$ and $r(t)$ denotes the regime index at time $t$, which takes values in $\{ 0, 1, \ldots, m \}$, and $\{ \epsilon_t \}$ is a stationary causal and invertible ARMA$(p,q)$ time series that applies to all regimes.  Observe that
\begin{align*}
\kappa_{t}=
     \begin{cases}
     \mu_0, \quad & 1 \le t \leq \tau_1,\\
     \mu_1, \quad & \tau_1+1 \le t \leq \tau_2,\\
     \; \;  \quad \vdots  \\
     \mu_m, \quad &\tau_{m}+1 \le t \leq N
     \end{cases}.
\end{align*}

There are many challenges in the multiple changepoint problem.  Here, estimation of a global autocovariance function that applies to all regimes --- considered further in Section \ref{sec:gac} --- is difficult.  One also has to estimate an unknown number of changepoints, their locations, and all segment parameters in a computationally feasible manner for some of the techniques.

While many authors have considered multiple changepoint issues, most assume IID $\{ \epsilon_t \}$.  For IID errors, dynamic programming based approaches \citep{AugerLawrence1989, killick2012}, model selection methods using LASSO \citep{harchaoui2010, shen2014}, and moving sum statistics \citep{kirch2014mosum} have all been applied to multiple changepoint problems --- this list is not exhaustive.  As in the AMOC setting, techniques for independent data may not work well for dependent series \citep{Davis_etal_2006, li_lund2012, chakar2017}. 

The multiple changepoint techniques considered here can be put into two broad categories: 1) recursive segmentation and algorithmic methods using AMOC techniques, and 2) direct approaches that fit all series subsegments jointly. The two approaches are completely different in their perspective.  Elaborating, recursive techniques employ AMOC single changepoint methods in an iterative manner, identifying at most one additional changepoint in each subsegment at each recursion level.  In contrast, direct techniques model and estimate the multiple changepoint configuration jointly; here, penalization methods typically drive the discourse. No hypothesis testing paradigm underlies any direct approach.  Some multiple changepoint techniques apply only to special time series structures. For example, \cite{chakar2017} is exclusively designed for AR$(1)$ series. Their techniques are not considered here as they cannot be applied to all of our considered scenarios.

\subsection{Recursive Segmentation}
Recursive segmentation approaches first focus on finding a single (usually the most prominent) changepoint, thereafter iterating in some manner to identify additional changepoints. The primary tool here is binary segmentation \citep{scott1974}, which provides a multiple changepoint configuration estimate via any AMOC method. Binary segmentation first tests the entire series for a single changepoint. Should a changepoint be found, the series is split about the changepoint time into two subsegments that are then analyzed for additional changepoints using the AMOC strategy. The process is repeated until no subsegment tests positive for a changepoint.  Binary segmentation works best when the changepoints are well separated and the segment means are distinct.  In our comparisons, the AMOC statistic adopted for recursive segmentation is the SCUSUM test applied to one-step-ahead prediction residuals, which won our AMOC comparisons in the previous section. 

Extensions of binary segmentation abound and include circular binary segmentation \citep{cbs2004}, which seeks to identify a segment of data that has a distinct mean from the rest of the series.  A popular binary segmentation extension considered here is wild binary segmentation (WBS) \cite{fryzlewicz2014}.  WBS samples subsegments of the entire data of varying lengths and performs an AMOC test on each sampled subsegment. \cite{fryzlewicz2014} suggests sampling at least $(9N^2)\log(N^2 \delta^{-1})/(\delta^2)$ subsegments, where $\delta$ is the minimum spacing between changepoints (see Assumption 3.2 of \cite{fryzlewicz2014}) as this produces a high probability of drawing a favorable subsegment.  WBS is a randomized search and hence may return different segmentations on different runs.  In our simulations, WBS employs a standard CUSUM test since its threshold was developed particularly for standard CUSUM methods. In addition, the threshold constant $C=1.3$ is used as suggested \citep{fryzlewicz2014}.

Binary segmentation approaches and their variants are simple to implement and are computationally fast. However, they are not guaranteed to achieve the global optimal solution as they essentially are a  ``greedy algorithm" that sequentially makes decisions based solely on information during the current step.  Also inherent in these approaches is the need for the AMOC statistic to behave appropriately when multiple changes are present --- this may not happen.

\subsection{Global Autocovariance Estimation}
\label{sec:gac}

For our work, the autocovariance of the series is assumed constant across time, applying to all series subsegments. This autocovariance function will be needed to decorrelate the series before applying any binary segmentation search methods to the one-step-ahead prediction residuals.  Unfortunately, accurate estimation of the autocovariance function requires knowledge of the underlying mean structure.  In the single changepoint case, the long-run covariance defined in \eqref{eq:lrcov} arises in the limit laws; however, this does not extend to multiple changepoint settings as no theoretical equivalent of \eqref{eq:lrteststat} exists.

In our setup, the second order (covariance) model parameters are deemed nuisance parameters and are estimated using the entire data sequence.  To account for the impact of unknown mean shifts on these estimators, Yule-Walker type moment equations will be used on the first order difference of $\{ X_t \}$.  The first order difference $X_t - X_{t-1}$ is used because $E[ X_t - X_{t-1} ]=0$ unless a changepoint occurs at time $t$. Define $d_t = X_t - X_{t-1}$ and note that $d_t = \epsilon_t - \epsilon_{t-1}$ except when time $t$ is a changepoint. Let $\gamma_d(h) =\mbox{Cov}(d_t, d_{t-h})$. For the AR($p$) case, which is our primary interest, estimators of the AR($p$) parameters formed from $\{ d_t \}$ take the form
\begin{equation}
\label{estimator}
\widehat{\boldsymbol{\phi}} =
\widehat{\boldsymbol{M}}^{-1}\widehat{\boldsymbol{\rho}}_d,
\end{equation}
where $\boldsymbol{\phi} = (\phi_1, \ldots, \phi_p)^\prime$ and 
\begin{equation*}    
\boldsymbol{M}=\begin{bmatrix}
\frac{1}{2} &-\frac{1}{2} &-\left(\frac{1}{2}+\rho_d(1)\right) &\cdots &-\left(\frac{1}{2}+\sum_{j=1}^{p-2}\rho_d(j)\right) \\
\rho_d(1) &\rho_d(0) &\rho_d(1) &\cdots &\rho_d(p-2)\\
\rho_d(2) &\rho_d(1) &\rho_d(0) &\cdots &\rho_d(p-3)\\
\vdots &\vdots  & \vdots &\ddots &\vdots\\
\rho_d(p-1) &\rho_d(p-2) &\rho_d(p-3) &\cdots &\rho_d(0)
\end{bmatrix}.
\end{equation*}

The elements in $\hat{\boldsymbol{M}}$ and $\hat{\boldsymbol{\rho}}_d$ simply replace $\rho_d(h)$ with
\[
\hat{\rho}_d(h)=\frac{\hat{\gamma}_d(h)}{\hat{\gamma}_d(0)}=
\frac{\sum_{t=2}^{n-h} (X_t-X_{t-1})(X_{t+h}-X_{t+h-1})}
{\sum_{t=2}^n (X_t-X_{t-1})^2}.
\]

While \cite{Shi_etal_2020} discuss these AR($p$) estimators in detail, the intuition behind them is that if $m$ is small relative to $N$, then the mean shifts will have negligible impact on the estimated covariance structure of the differences since $X_t - X_{t-1} = \epsilon_t - \epsilon_{t-1}$ except at times $t$ that are changepoint times. \cite{Shi_etal_2020} demonstrate that this estimate of the covariance outperforms alternatives such as direct and windowed estimation.  Due to this, the Yule-Walker moment estimators in (\ref{estimator}) will be used in our simulations to decorrelate the series.

\subsection{Direct Modelling}
Direct modelling approaches analyze the whole series at once, optimizing an objective function with a penalty term that controls the number of changepoints. The techniques seek a changepoint configuration that minimizes
\begin{equation}
F(m; \tau_1, \ldots, \tau_m) := C(m;\tau_1,\ldots,\tau_m) + P(m;\tau_1,\ldots,\tau_m),
\label{PenLKHD}
\end{equation}
where $C$ is the cost of putting $m$ changepoints at the times $\tau_1, \ldots, \tau_m$ and $P$ is a penalty term to prevent over-fitting. There are many ways to define the cost and penalties. A frequently used cost is the negative log-likelihood. Here, we will use
\[
C(m;\tau_1, \ldots, \tau_m) = 
-2 \log(L_{\rm opt}(\boldsymbol{\theta}|m; \tau_1, \ldots \tau_m)),
\]
where $L_{\rm opt}(\boldsymbol{\theta}|m; \tau_1, \ldots, \tau_m)$ is the time series likelihood (Gaussian based) optimized over all parameters $\boldsymbol{\theta}$ given that $m$ changepoints occur at the times $\tau_1, \ldots, \tau_m$.

Penalties can be constructed in a variety of ways. Common penalties include minimum description lengths (MDL), modified Bayesian Information Criterion (mBIC), and the classic BIC penalty. AIC is another popular penalty, despite it not providing consistent estimates of the number or locations of the changepoint(s).  Of these four penalties, AIC and BIC are simple multiples of the number of changepoints, while the MDL and mBIC further incorporate changepoint time information.  The form of these penalties are listed in the following table.

\begin{table}[H]
\caption{Penalized Likelihood Objective Functions}
\begin{center}
\begin{tabular}{ c | l  }
Criteria    &   Objective Function \\
\hline\\ [-0.5em]
AIC      & $N\ln(\hat{\sigma}^2) + 2(2m+3)$     \\
[0.5em]
BIC      & $N\ln(\hat{\sigma}^2) + (2m+2)\ln(N)$     \\
[0.5em]
mBIC     & $\frac{N}{2}\ln(\hat{\sigma}^2) + \frac{3}{2}m\ln(N)+\frac{1}{2}\sum_{i=1}^{m+1} \ln \left( \frac{\tau_i - \tau_{i-1}}{N} \right)$      \\
[0.5em]
MDL & $\frac{N}{2}\ln(\hat{\sigma}^2) + \ln(m) + \frac{1}{2}\sum_{i=1}^{m+1} \ln (\tau_i - \tau_{i-1}) + \sum_{i=2}^m \ln(\tau_i)$ \\
[0.5em]
\hline
\end{tabular}
\end{center}
\end{table}
\noindent Here, $\hat{\sigma}^2$ is the estimated white noise variance of the $\{ \epsilon_t \}$ process that drives the ARMA errors.

MDL penalties are based on information theory and are discussed further in \cite{Davis_etal_2006} and \cite{li_lund2012}. The mBIC penalty is developed in \cite{zhang2007mbic}.  These two penalties are taken as zero when $m=0$.  The mBIC penalty tends to be larger for the same changepoint configuration than the MDL penalty; as such, MDL will often select models with more changepoints than mBIC. 

With penalized likelihood approaches, a computational bottleneck arises. Since there are $\binom{N-1}{m}$ different admissible changepoint configurations in a series with $m$ changepoints (time $N$ cannot be a changepoint), there are $2^{N-1}$ different changepoint configurations to consider when analyzing the entire series. This huge count makes an exhaustive model search --- one that evaluates all admissible changepoint configurations --- virtually impossible to conduct, even when $N$ is a small as 100. Unfortunately, PELT \citep{killick2012} and FPOP \citep{fpop}, two rapid dynamic programming based techniques, require the objective function to be additive over distinct regimes.  The presence of global parameters like the autocovariance function violates this restriction. Regime-additive likelihoods will not arise when $\{ \epsilon_t \}$ is ARMA($p,q$), although \cite{BaiPerron1998} argues that any boundary contribution is negligible if the ARMA parameters are allowed to change at each changepoint time (this is not the case here). Instead, a genetic algorithm will be used to optimize our penalized likelihoods exactly (including global autocovariances).

Elaborating on the above, optimal changepoint configurations estimated by penalized likelihoods can be found through a genetic algorithm (GA) search \citep{Davis_etal_2006, li_lund2012}. A GA is an intelligent random walk search that is unlikely to evaluate suboptimal changepoint configurations.  Unfortunately, the objective function in (\ref{PenLKHD}) is not convex, and its optimization is problematic; additional research is needed to optimize these penalized likelihoods.  Our GA encodes the changepoint configuration into a binary string and uses the \texttt{R GA} package from \cite{scrucca2013ga}. This GA has worked reliably in our experience.  

\section{Multiple Changepoint Simulations}
\label{sec:multsim}

In presenting simulation results for different scenarios, we will only present graphic(s) that are judged informative.  In general, for each simulation case considered, graphics of distances, average number of detected changepoints, and empirical probabilities of estimating the correct number of changepoints were produced.  The supplementary material contains any graphics that are not included here.  Similarly, we focus on unit shift mean sizes in the main text body unless otherwise noted; results for different shift sizes are presented in the supplementary material.

\begin{figure}[H]
\centering
\includegraphics[scale=1.2]{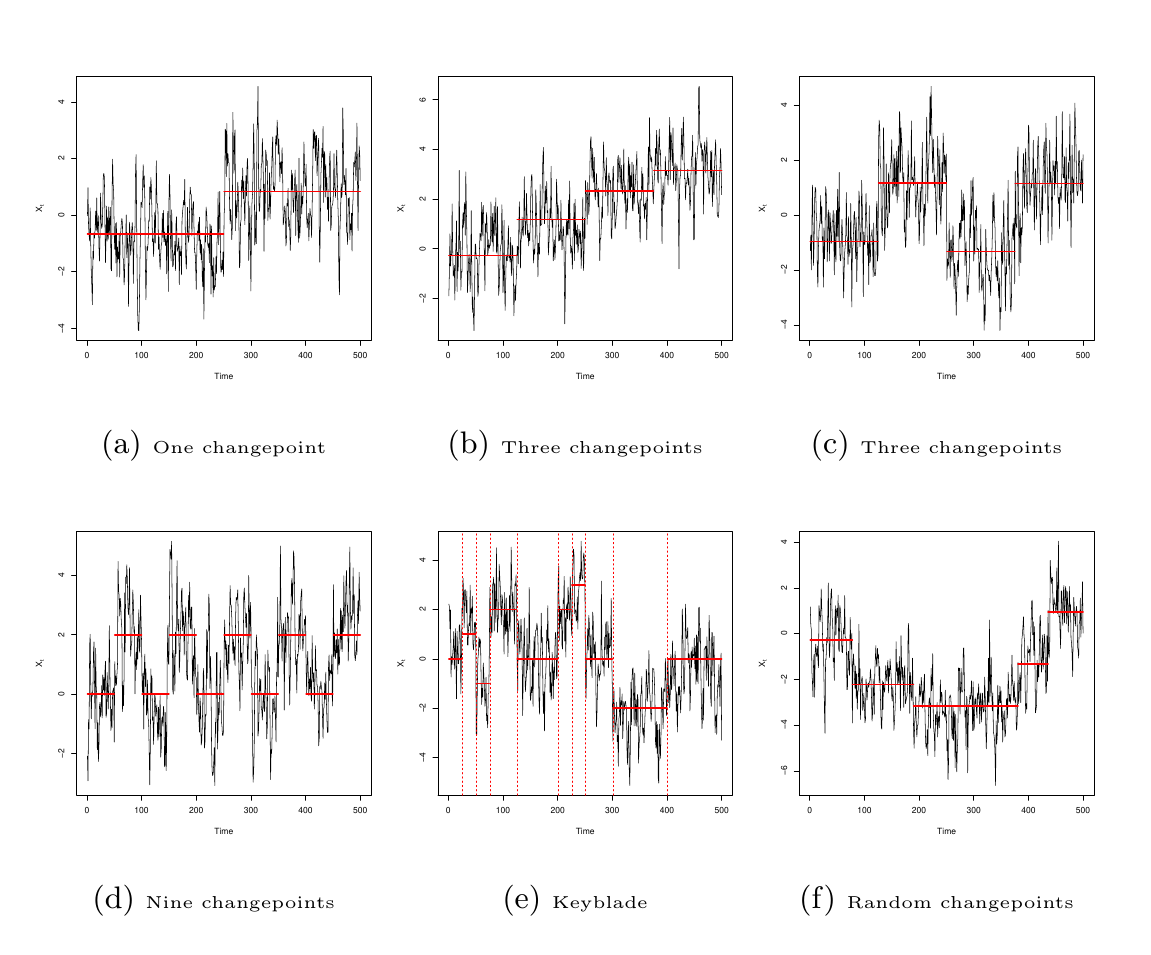}
\caption{AR(1) Time Series with Different Changepoint Settings}
\label{fig:cpt_demo} 
\end{figure}

The changepoint configurations that we consider are illustrated in Figure \ref{fig:cpt_demo}, which shows sample time series generated with the various mean shift configurations. These configurations range from scenarios with no or few changepoints to those with a large number of changepoints.  All series have length $N=500$.  

AIC performs miserably in all our scenarios, always selecting an excessive number of changepoints.  Since plotting AIC results would distort all other graphical comparisons, AIC results are not presented to accentuate differences between the remaining methods.

\subsection{Comparing Multiple Changepoint Segmentations}
\label{metric}
Before presenting our simulations, we discuss how to compare an estimated multiple changepoint segmentation to its true value.  The estimated multiple changepoint configuration could have a different number of changepoints than the true configuration.   For a single changepoint method, such a comparison is easy:  examine first whether the method flags a changepoint, and then any distance from the true changepoint time.  With multiple changepoint configurations, this comparison is complicated by the fact that different segmentations may have different numbers of changepoints:  which changepoint times in one particular configuration correspond to those in another may be nebulous.

To compare different methods, a distance between the two changepoint configurations $\mathcal {C}_1= (m;\tau_1, \ldots, \tau_m)$ and $\mathcal {C}_2=(k;\eta_1, \ldots, \eta_k)$ will now be developed. Several distances have been utilized by the multiple changepoint field.  Some, such as the mean squared error (MSE) of the fitted means, V-measure, or Hausdorff distance, are not specific to changepoint problems.   Others, such as the number of changepoints or true/false positive detection rates, are more tailored to the changepoint problem.  However, each of these statistics quantifies only one aspect of the fit.  For example, the MSE could be low, but the number of changepoints could still be overestimated; or the number of changepoints could be perfect, but their locations could be inaccurate.  As such, we introduce a new changepoint-specific distance balancing the two key components of multiple changepoint analysis:  1) the number of changepoints and 2) their individual locations.

To balance the number and location aspects of changepoint configurations, two components in our distance are needed.  The first measures the discrepancy in the numbers of changepoints in the two configurations, for which we use absolute difference.  The second component measures discrepancies in the changepoint times.  This is trickier to quantify as the number of changepoints may be different in the two configurations and some sort of ``matching procedure" is needed. For two changepoint segmentations, $\mathcal{C}_1$ and $\mathcal{C}_2$, the distance used here is
\begin{equation}      
    d(\mathcal{C}_1, \mathcal{C}_2) = |m-k|+ \min \{ \mathcal{A}(\mathcal{C}_1, \mathcal{C}_2 ) \}. \label{eqn:cpt_dist}
\end{equation}
The term $|m-k|$ assigns the difference in changepoint numbers for any mismatch in the total number of changepoints. The term $\min \mathcal{A} (\mathcal{C}_1, \mathcal{C}_2)$ reflects the smallest cost that matches changepoint locations in $\mathcal{C}_1$ to those in $\mathcal{C}_2$.  This term can be computed via the following linear assignment methods:
\begin{equation*}
\mathcal{A} ({\cal C}_1, {\cal C}_2)=
\sum_{i=1}^k\sum_{j=1}^m c_{i,j}I_{i,j}
\end{equation*}
subject to the constraints $\sum_{i=1}^k I_{i,j}=1$, for $j \in \{ 1, \ldots, m\}$ and $\sum_{j=1}^m I_{i,j} \leq 1$ for $i \in \{ 1, \ldots, k \}$.  Here, the cost of assigning $\tau_i$ to $\eta_j$ is taken simply as $c_{i,j} = |\tau_i-\eta_j|/N$ and $I_{i,j} \in \{ 0,1 \}$ is the decision variable
\begin{equation*}
I_{i,j} = 
\begin{cases}
\; 1 \qquad &\text{if $\tau_i$ is assigned to $\eta_j$}\\
\; 0 \qquad &\text{otherwise}
\end{cases}.
\end{equation*}
This linear assignment problem can be efficiently computed from algorithms in \cite{burkard2012assignment}.

One can verify that (\ref{eqn:cpt_dist}) defines a legitimate distance satisfying the triangle inequality. The larger the distance is, the worse the two configurations correspond to one another.  The term $\min \mathcal{A}(\mathcal{C}_1, \mathcal{C}_2)$ can be shown to be bounded by unity and measures how closely the two changepoint configurations match up to one another.   When both configurations have many changepoints, the distance is dominated by the $|m-k|$ term.  In our simulations, estimated multiple changepoint configurations will be compared to the true changepoint configuration with this distance.  

\subsection{No Changepoints}

Many modern multiple changepoint simulation studies increasingly focusing on cases with a large number of changepoints, eschewing single and no changepoint scenarios.  We include such scenarios here to help illuminate the differences between the methods.

\begin{figure}[H]
\centering
\includegraphics[scale=0.5]{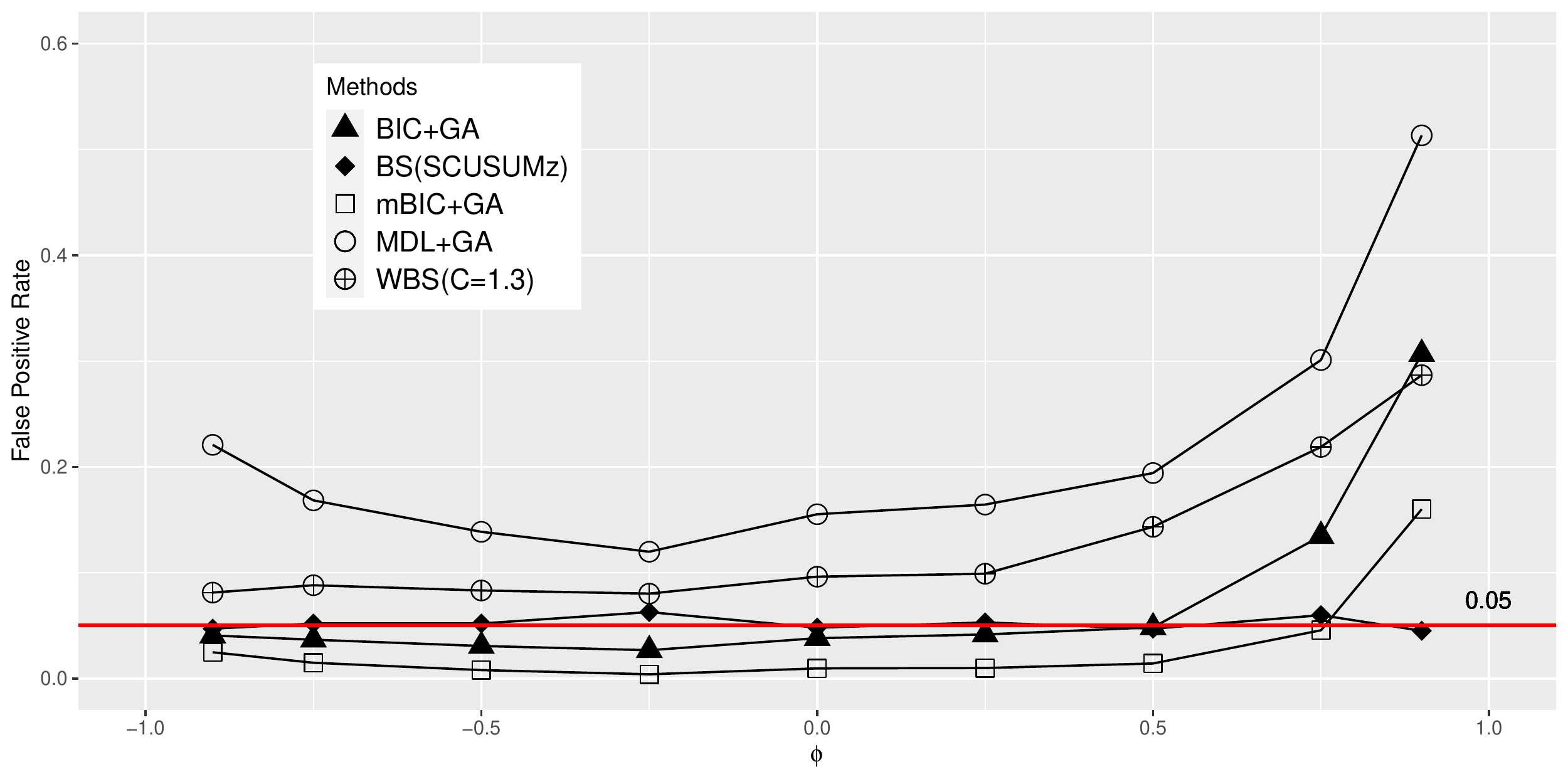}
\caption{Empirical False Positive Detection Rates for an AR$(1)$ Series with Various $\phi$.  Truth: No Changepoints.}
\label{fig:ar1_0mcpt_fprate} 
\end{figure}

Our first simulation considers the changepoint free case in an AR(1) Gaussian series having various correlation parameters $\phi$, $N=500$, and $\sigma^2=1$.  Figure \ref{fig:ar1_0mcpt_fprate} shows probabilities of falsely declaring one or more changepoints over 1,000 independent simulations.  Unlike the single changepoint case, the methods here are not aiming to control any false positive rate.

The results show that BIC, mBIC, and binary segmentation perform best, with WBS and MDL performing significantly worse. It is worth noting that WBS has a signicantly higher false positive rate, an issue discussed further in \cite{lund2020wbs2}. Binary segmentation is arguably best here, an expected finding since there are no changepoints (an AMOC test applied to the series' one-step-ahead prediction residuals should not see a changepoint and stop any recursion at its onset).  All methods perform better with negative $\phi$ than with positive $\phi$; performance of all methods degrades as $\phi$ moves upwards towards unity (as expected).

\subsection{A Single Changepoint in the Record's Middle}
\begin{figure}[H]
\centering
\includegraphics[scale=0.5]{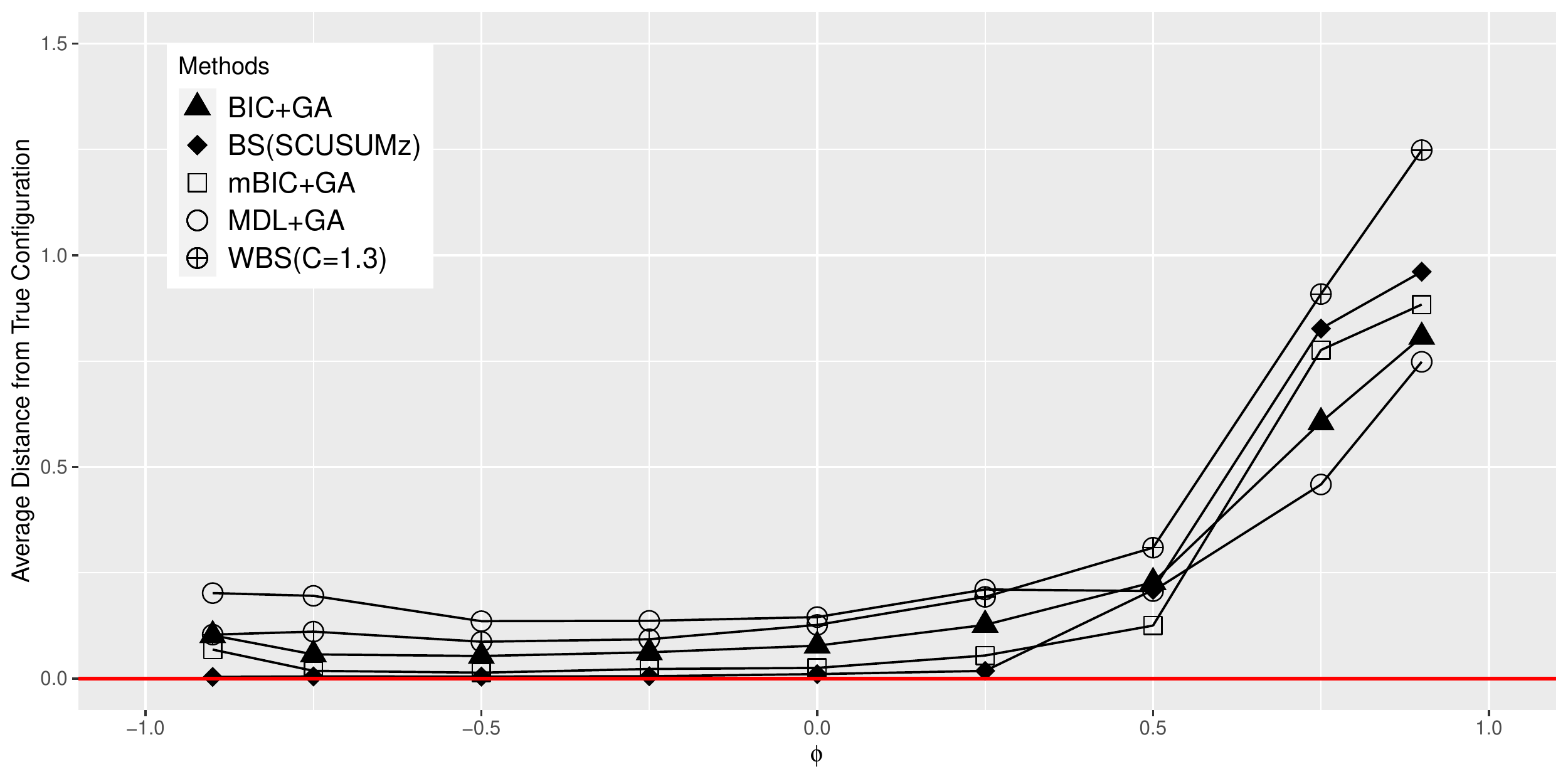}
\caption{Average Distances for an AR$(1)$ Series with Varying $\phi$.  Truth:  One Changepoint in the Middle Moving the Series Upwards.}
\label{fig:ar1_1mcpt_dist}
\end{figure}

We now move to simulations with one changepoint in the same AR(1) setup above.  The changepoint is placed in the middle of the series, $t=251$. Figure \ref{fig:ar1_1mcpt_dist} shows the average distances between the estimated changepoint configurations and the true configuration.  While there are no huge discrepancies between the methods, for heavily correlated series, binary segmentation is the worst and MDL and mBIC the best.  Again, all tests degrade as $\phi$ approaches unity.  MDL has the least variability across $\phi$.  Comparing to the single changepoint results, the multiple changepoint penalties are more conservative than the LRT.  Also, since the average distance is less than unity, the correct number of changepoints is often being identified.

\subsection{A Three Changepoint Staircase}
\begin{figure}[H]
\centering
\includegraphics[scale=0.5]{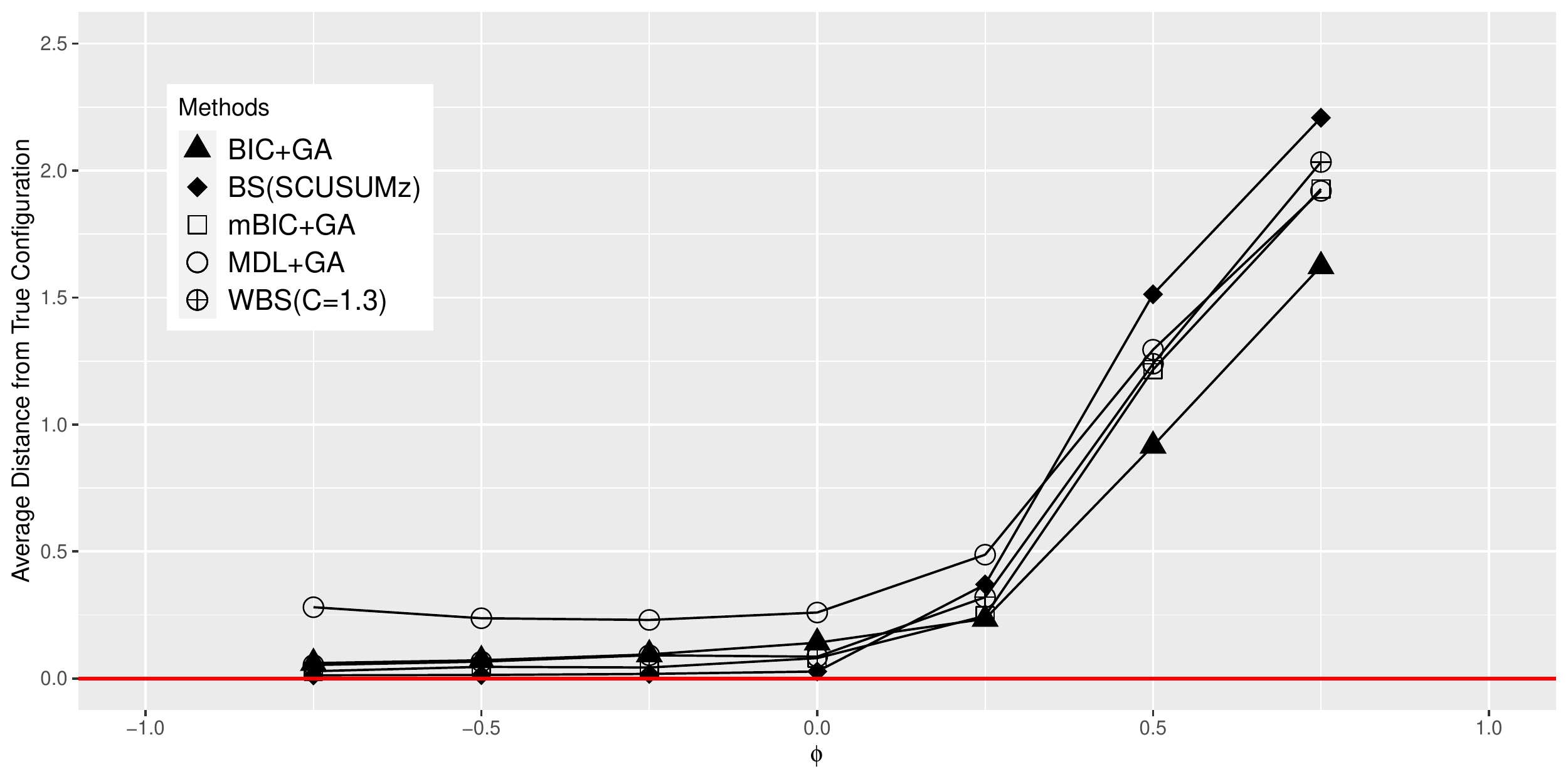}
\caption{Average Distances for AR$(1)$ Series with Different $\phi$. Truth:  Three Equally Spaced Changepoints Moving the Series Up-Up-Up.}
\label{fig:ar1_3mcpt_uuu_dist}
\end{figure}

Our next case moves to a setting with three mean shifts, partitioning the series into four equal-length regimes. The changepoints occur at times 126, 251, and 376, with each changepoint shifting the series upward by one unit (up-up-up). As before, Figure \ref{fig:ar1_3mcpt_uuu_dist} reports average distances. MDL performs the worst for negative $\phi$, while the other methods perform similarly. Perhaps surprisingly, binary segmentation starts to degrade when $\phi$ becomes positive, with the others also degrading, but to a lesser extent.  BIC performs best across all $\phi$.  

\subsection{Three Alternating Changepoints}
\begin{figure}[H]
\centering
\includegraphics[scale=0.5]{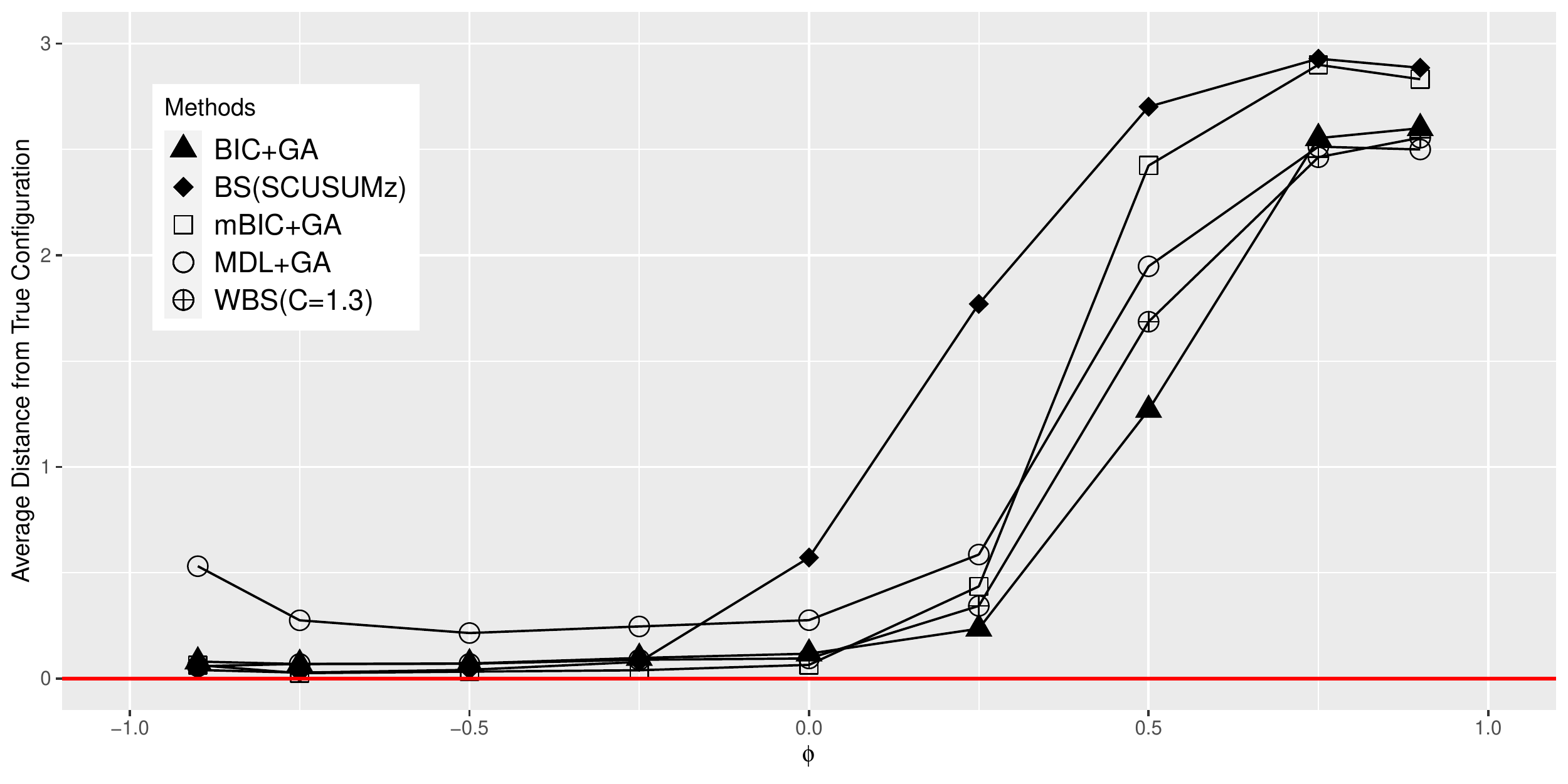}
\caption{Average Distances for an AR$(1)$ Series with Varying $\phi$. Truth:   Three Equally Spaced Changepoints Moving the Series Up-Down-Up.}
\label{fig:ar1_3mcpt_dist}
\end{figure}

Next, we consider another three changepoint configuration, the changepoint times again being equally spaced, but this time moving the series up, then down, and then up again (up-down-up). Figure  \ref{fig:ar1_3mcpt_dist} reports the distances.  All methods have a harder time than with the last up-up-up changepoint configuration.  In this setting, binary segmentation becomes fooled and estimates too few changepoints; mBIC is also not doing as well as the other methods.  MDL and WBS work better, the surprise winner being BIC.

\subsection{A Nine Changepoint Staircase}
\begin{figure}[H]
\centering
\includegraphics[scale=0.5]{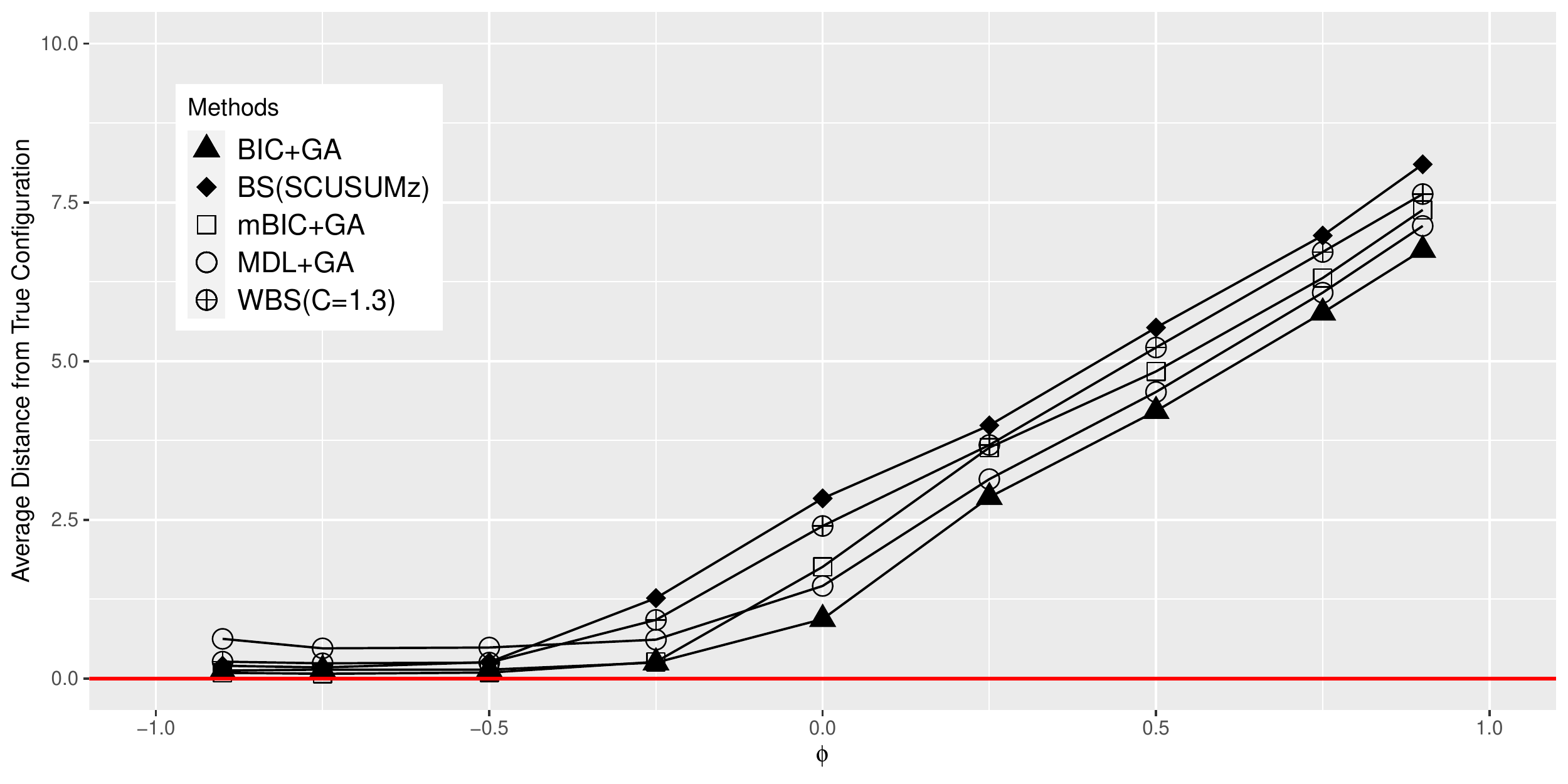}
\caption{Average Distances for an AR$(1)$ Series with Varying $\phi$. Truth: Nine Changepoints, All Up.}
\label{fig:ar1_9mcpt_uuu_dist}
\end{figure}

Next, we move to cases with nine changepoints.  Our first set of simulations equally spaces all changepoint times in the record, each moving the series higher (All Up).  Because the changepoints are more difficult to detect, we have increased the absolute mean shift magnitude to two units --- this induces more separation between the methods, allowing for an easier comparison. Figure \ref {fig:ar1_9mcpt_uuu_dist} displays distances for this setting. The winners are BIC and MDL; losers are WBS and binary segmentation. 

\subsection{Nine Alternating Changepoints}
\begin{figure}[H]
\centering
\includegraphics[scale=0.5]{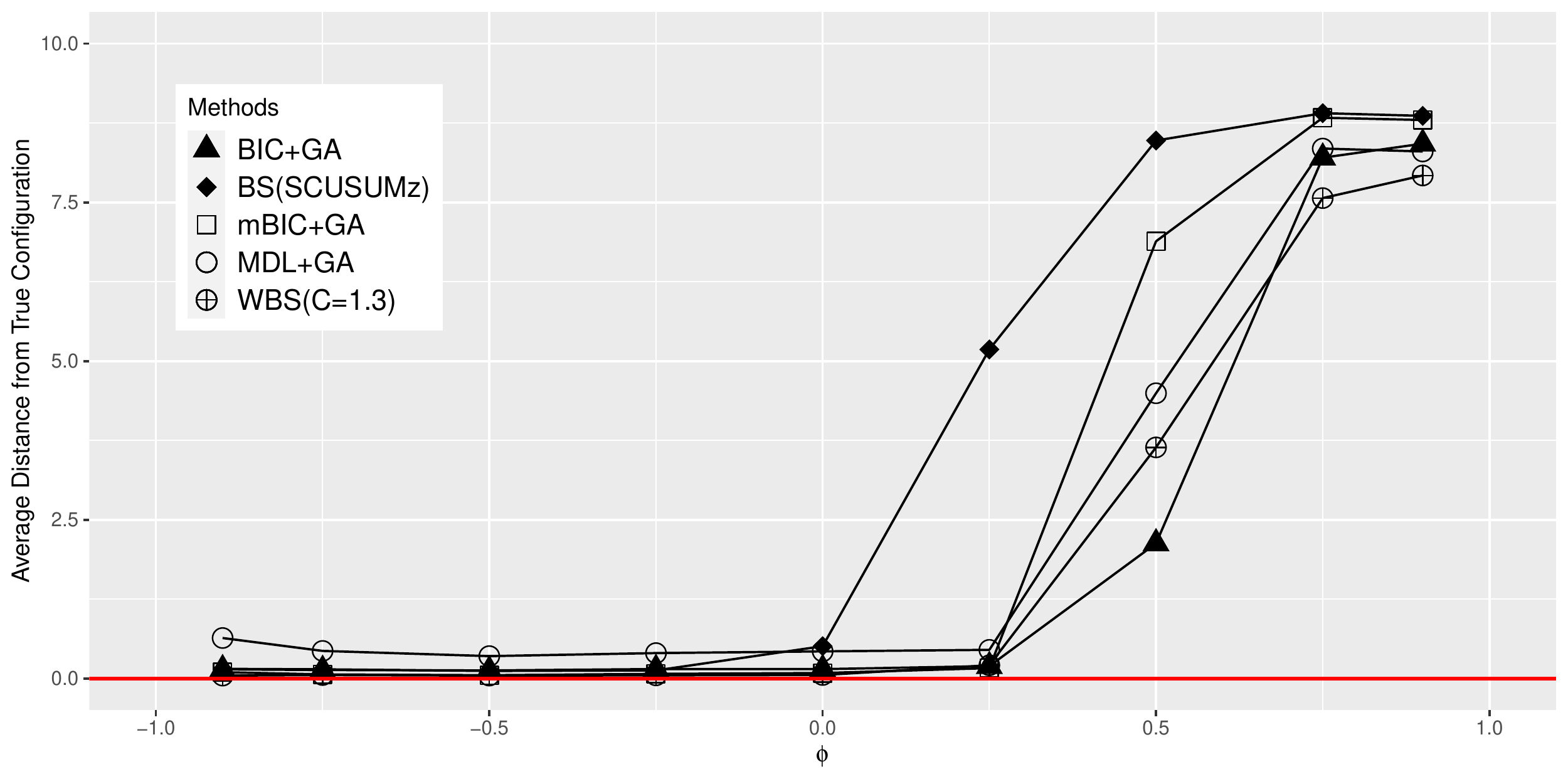}
\caption{Average Distances for an AR$(1)$ Series with Varying $\phi$. Truth: Nine Alternating Changepoints.}
\label{fig:ar1_9mcpt_Delta=2_dist}
\end{figure}

Our next set of simulations again considers nine changepoints, but the directions of the equally spaced mean shift sizes of magnitude two are now alternated in an Up-Down-Up-Down-Up-Down-Up-Down-Up fashion (alternating).  Figure \ref{fig:ar1_9mcpt_Delta=2_dist} displays results.  The best method here is BIC again with WBS doing better than in the previous setting; mBIC is a laggard and binary segmentation is again the worst.

\subsection{Nine Keyblade Changepoints}
\begin{figure}[H]
\centering
\includegraphics[scale=0.6]{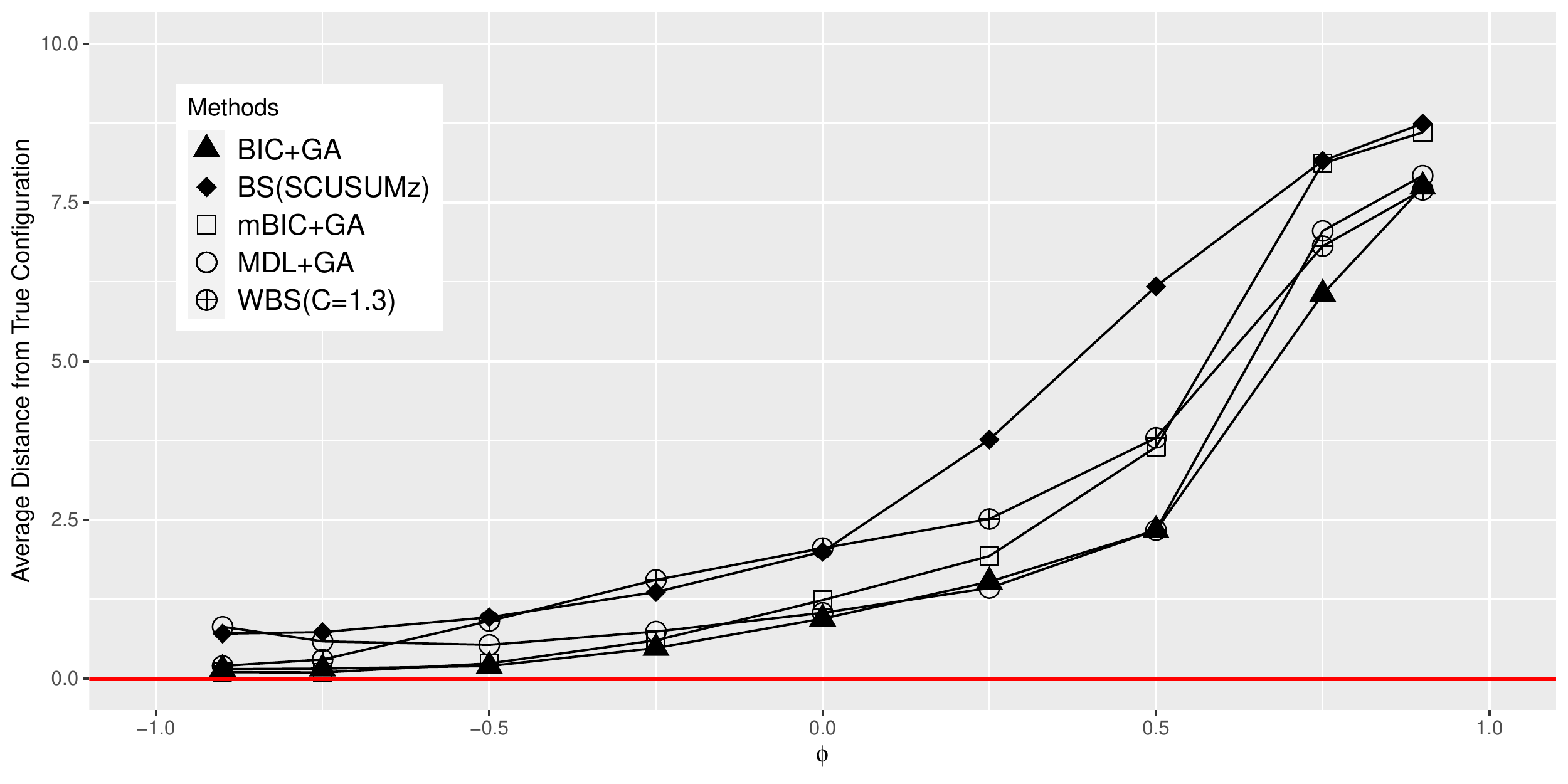}
\caption{Average Distances for the Keyblade AR$(1)$ Series with Varying $\phi$. Truth: Nine Changepoints.}
\label{fig:keybalde_dist}
\end{figure}

As a different type of setup, we consider the nine changepoint setting where the sizes of the nine mean shifts vary, their shift directions vary, and the changepoint times are not equally spaced. Figure \ref{fig:cpt_demo}(d) shows our chosen pattern for $E[X_t]$, which we call a ``keyblade". The distances in Figure \ref{fig:keybalde_dist} reveal BIC and MDL as winners, and WBS and binary segmentation as inferior.

\subsection{Random Changepoints}
\begin{figure}[H]
\centering
\includegraphics[scale=0.5]{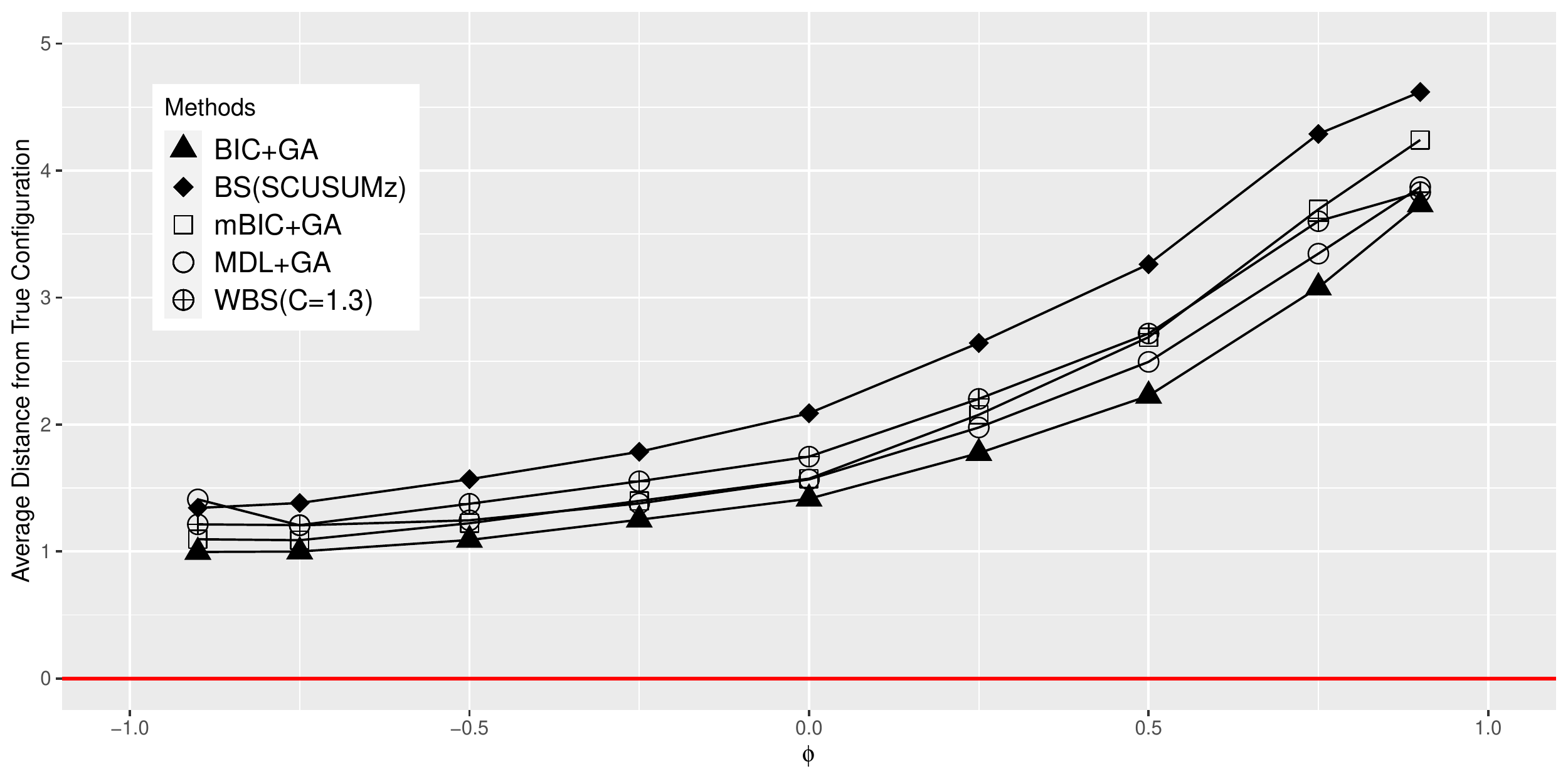}
\caption{Average Distance between the Estimated and True Changepoint Locations.}
\label{fig:rand_dist}
\end{figure}

We now consider settings with a random number of changepoints simulated from a Poisson distribution with a mean of five. The locations of any mean shifts are placed uniformly in the set $\{ 2, \ldots, N \}$ without replacement. The mean of each segment is simulated from a normal distribution with a zero mean and a standard deviation of $1.5$. Figures \ref{fig:rand_dist} summarizes the results: BIC and MDL are again superior and binary segmentation inferior. 

\subsection{Varying Series Lengths}
It was surprising to us how well the simple BIC penalty has done so far --- especially since this penalty does not depend on the changepoint times.  To examine this issue further, we fix the AR(1) parameter at $\phi=0.5$ and compare BIC and mBIC distances as $N$ varies with one and three changepoints.  Here, the changepoints induce equal length regimes, all mean shift sizes are of a unit magnitude, and their directions alternate with the first direction being upwards. Table \ref{tab:vary_N} reports average BIC and mBIC distances when $N \in \{ 500, 1000, 2500 \}$.  As the sample size increases, the additional penalty the mBIC has on the length of the segments results in fewer changepoints identified than for BIC. As $N$ grows, there is a tendency for BIC to add (erroneous) changepoints in some samples.  Thus, as the number of changepoints and $N$ grows, mBIC is better than BIC.  This leads us to recommend mBIC over BIC for larger $N$ or numbers of changepoints.

\begin{center}
\begin{table}[H]
\centering
\caption{Comparison of BIC and mBIC. Truth: $m$ changepoints, all of a unit magnituide, placed in alternating directions that equally space the record length for an AR$(1)$ series with varying lengths $N$. Here, $\sigma^2=1$ and $\phi=0.5$.}
  \begin{tabular}{lSSSS}
    \toprule
    \multirow{2}{*}{Avg.~Distance} &
      \multicolumn{2}{c}{$m=1$}   &
      \multicolumn{2}{c}{$m=3$} \\
      & BIC & mBIC & BIC & mBIC \\
      \midrule
    $N=500$  & 0.227 & 0.125 & 1.270 & 2.420  \\
    $N=1000$ & 0.126 & 0.066 & 0.311 & 0.921 \\
    $N=2500$ & 0.121 & 0.047 & 0.123 & 0.066 \\
    \bottomrule
  \end{tabular}
  \label{tab:vary_N}
\end{table}
\end{center}

Before moving to non-AR(1) settings, we examine method performance as the mean shift magnitudes increase. Here, we fix $N=500, \phi=0.5$, and $\sigma^2=1$ and consider three alternating changepoints placed at the times $126, 251$, and $376$.   Mean shift magnitudes $\Delta$ are varied from 1 to 3. Average distances over $1,000$ simulations are reported in Table \ref{tab:different_Delta}. As the mean shift magnitudes increases, all methods improve.  BIC and MDL, two frequent winners of past scenarios, perform worst when the mean shift size is largest; moreover, WBS and binary segmentation, two frequent past losers, perform best.   mBIC reports the smallest average distance when $\Delta \geq 2$.  

\begin{table}[H]
\centering
\footnotesize{ 
\begin{tabular}{ c|cccccc } 
$\Delta$	&BIC+GA		&mBIC+GA	&MDL+GA	&BS(SCUSUMz) &WBS(C=1.3)\\
\hline
\hline
\\[-0.6em]
$\Delta=1$ & 1.269	&2.424	&1.948	&2.702	&1.686
\\
\\[-0.6em]
$\Delta=2$ &0.140	&0.051	&0.209	&0.843	&0.149
\\
\\[-0.6em]
$\Delta=3$ & 0.126	&0.042	&0.188	&0.077	&0.079
\\
\\[-0.6em]
\hline
\end{tabular}
}
\caption{Average Distance for an AR$(1)$ Series with Varying Mean Shift Magnitudes}
    \label{tab:different_Delta}
\end{table}

Our final simulation task considers other autoregressive error structures.  We begin with AR(2) errors and the case of no changepoints.  Table \ref{tab:ar2_0mcpt_fprate} shows false positive rates of signaling one or more changepoints when in truth none exist for various AR(2) parameters $\phi_1$ and $\phi_2$.  In this and all four tables below, 1,000 independent simulations are conducted, $N=500$, $\sigma^2=1$, and all mean shift sizes are two units (this adds additional information to cases above where mean shifts were of a unit size).   All four tables are discussed below in tandem after they are presented.

\begin{table}[H]
\centering
\footnotesize{ 
\begin{tabular}{ c|cccccc } 
$\{\phi_1, \phi_2\}$	&BIC+GA		&mBIC+GA	&MDL+GA	&BS(SCUSUMz) &WBS(C=1.3)\\
\hline
\hline
\{0.6, 0.35\} & 21.5\% &2.5\%
&38.8\% & 22.6\% & 50.0\%
 \\
\{0.6, 0.3\} &17.5\% &2.6\% &33.2\% &10.2\% &36.6\% 
\\
\{0.6, -0.1\} & 5.9\% &1.1\% &15.6\% &0.3\% & 17.4\%
\\
\{0.5, -0.2\} & 4.1\%
&1.6\%
&13.6\%
&0.0\%
&11.7\%
\\
\{0.2, -0.5\} & 3.0\%
&0.6\%
&9.4\%
&0.1\%
&9.1\%
\\
\hline
\end{tabular}
}
\caption{False Positive Rates for an AR$(2)$ Series with Varying $\{\phi_1, \phi_2\}$. Truth: No Changepoints}
    \label{tab:ar2_0mcpt_fprate}
\end{table}

Table \ref{tab:ar2_3mcpt_dist} reports average distances for the AR(2) scenario of the last table, but now with three changepoints. The three shifts induce four equal length regimes and shift the series mean in an up-down-up manner.

\begin{table}[H]
    \centering
\footnotesize{ 
\begin{tabular}{ c|cccccc } 
$\{\phi_1, \phi_2\}$	&BIC+GA		&mBIC+GA	&MDL+GA	&BS(SCUSUMz) &WBS(C=1.3)\\
\hline
\hline
\{0.6, 0.35\} & 2.757 & 2.932 &2.759 &2.633 &2.265 \\
\{0.6, 0.30\} & 2.484 & 2.895 &2.510 &2.742 &2.337 \\
\{0.6, -0.1\} & 0.167 & 0.052 &0.182 &0.818 &0.193 \\
\{0.5, -0.2\} & 0.131 & 0.032 &0.163 &0.072 &0.101 \\
\{0.2, -0.5\} & 0.086 & 0.023 &0.111 &0.040 &0.068 \\
\hline
\end{tabular}
}
\caption{Average Distances for an AR$(2)$ Series with Varying $\{\phi_1, \phi_2\}$. Truth: Three Alternating Changepoints of Size $\Delta=2$.}
    \label{tab:ar2_3mcpt_dist}
\end{table}

Table \ref{tab:ar4_0mcpt_fprate} shows false positive rates of signaling one or more changepoints when in truth there are none for various parameter choices in an AR(4) series.

\begin{table}[H]
\centering
\footnotesize{ 
\begin{tabular}{ c|cccccc } 
$\{\phi_1,\; \phi_2,\; \phi_3,\; \phi_4 \}$	&BIC+GA		&mBIC+GA	&MDL+GA	&BS(SCUSUMz) &WBS(C=1.3)\\
\hline
\hline
\{0.5,\;0.25,\;0.15,\;0.05\} &66.5 \%
&44.8\% 
&76.4\%
&29.7\%
&54.4\% 
\\
\{0.6,\; 0.3,\; 0.1, \; -0.3\} &16.7 \%
&8.5\%
&42.0\%
&0.6\% 
&21.5\%
\\
\{0.6,\;0.3,\; -0.3,\; -0.1\} & 9.9\%
&4.9\%
&32.5\%
&0.1\%
&14.8\%
\\
\{0.6,\;-0.4,\; -0.2,\; -0.1\} &5.0\%
&2.5\%
&27.0\% 
&0.2\%
&10.3\%
\\
\{0.6,\; -0.4,\; 0.3,\; -0.2\} & 5.3\%
&1.6\% 
&22.9\%
&0.2\%
&17.4\%
\\
\hline
\end{tabular}
}
\caption{False Positive Rates for an AR$(4)$ Series with Varying $\{\phi_1, \phi_2,\phi_3, \phi_4\}$. Truth: No Changepoints}
\label{tab:ar4_0mcpt_fprate}
\end{table}

Finally, Table \ref{tab:ar4_3mcpt_dist} reports average distances over 1,000 independent simulations for the same AR(4) scenario above.  The mean shift specifications are repeated from Table \ref{tab:ar2_3mcpt_dist}.

\begin{table}[H]
\centering
\footnotesize{ 
\begin{tabular}{ c|cccccc } 
$\{\phi_1,\; \phi_2,\; \phi_3,\; \phi_4 \}$	&BIC+GA		&mBIC+GA	&MDL+GA	&BS(SCUSUMz) &WBS(C=1.3)\\
\hline
\hline
\{0.5,\; 0.25,\;0.15,\;0.05\} 
&2.723
&2.420
&3.360
&2.516
&2.151
\\
\{0.6,\; 0.3,\; 0.1, \; -0.3\} &0.615
&1.582
&1.292
&2.318
&1.256
\\
\{0.6,\; 0.3,\; -0.3,\; -0.1\}  &0.205
&0.107
&0.251
&0.834
&0.211
\\
\{0.6,\; -0.4,\; -0.2,\; -0.1\} &0.127
&0.055
&0.319
&0.031
&0.079
\\
\{0.6,\; -0.4,\; 0.3,\; -0.2\}
& 0.161
&0.066
&0.246
&0.228
&0.101
\\
\hline
\end{tabular}
}
\caption{Average Distances for AR$(4)$ Errors with Varying $\{\phi_1, \phi_2, \phi_3, \phi_4\}$. Truth: Three Alternating Changepoints of Size $\Delta=2$.}
\label{tab:ar4_3mcpt_dist}
\end{table}

In the above tables, when there are no changepoints, binary segmentation appears best and MDL and WBS worst, repeating conclusions for AR(1) errors.  In the tables with three changepoints and heavily positively correlated errors, MDL, BIC, and WBS all do comparatively well; when the correlation becomes negative, the situation reverses and mBIC and binary segmentation are best.   These aspects were also seen for AR(1) series, although we did not remark about the negatively correlated results.

To summarize our overall conclusions on multiple changepoints, the following points emerge:

\begin{itemize}
\item{AIC and binary segmentation are not competitive.   Binary segmentation worked well only when no or few changepoints existed and worsened when multiple mean shifts act in opposite directions.  We do not recommend either of these techniques.}

\item{Although not depending on the changepoint configuration, BIC is surprisingly good across a wide range of scenarios.  However, as $N$ gets larger, mBIC becomes superior.}

\item{MDL was often the best penalized likelihood technique in heavily correlated scenarios, but does not work as well with negatively correlated series.  MDL also tends to lose to mBIC when the changepoint mean shift sizes are large or when changepoints are infrequent.}

\item{MDL and WBS techniques should be used with caution if there is a possibility that no changes are present, as they have high false positive rates.}

\item{BIC and mBIC perform well in the low frequency changepoint settings.}

\end{itemize}
We close with one more comment that is not apparent from the reported results.  The MDL penalty works reasonably in a large variety of positively correlated scenarios.  However, when it is wrong, it has a tendency to put several changepoints times very closely to one and other, typically near the beginning of the record.   This is an attempt by the method to flag an outlier. If one imposes a minimum spacing between changepoint times to combat this, the method becomes much better.

\section{Comments and Conclusions}
\label{sec:conc}
This paper presented a systematic comparison of common single and multiple changepoint techniques in time series settings.   Previous work had demonstrated how blindly applying techniques that assume IID to data could lead to erroneous conclusions.  Here, we focused on how IID methods could be modified for the time series setting, either by correcting the asymptotic distribution, or by modifying the test statistic.

In constructing our comprehensive approach, a summary of the major different techniques available was made in a single manuscript; hence, this paper has utility as a reference.  A new changepoint distance was also developed that combines the two important features of changepoint detection, identification of the correct number and location(s) of the changepoints, within a single metric.

In the single changepoint case, it was found that the best techniques apply IID methods to the time series of one-step-ahead prediction residuals. The best performing single changepoint detection method was the sum of CUSUM statistic in \cite{bai1993partial}.  Extreme value based asymptotic tests exhibited poor detection power.  

In the multiple changepoint case, conclusions were more nebulous; however, binary segmentation and AIC are not recommended.   The penalized likelihoods MDL, mBIC, and BIC all are worthy of additional study. WBS methods also performed reasonably and deserve additional attention, especially given their relatively recent entrance into the literature.  At this point, it is still not clear whether pure algorithmic techniques can beat penalized likelihood methods.  It is our view that one should use BIC penalized likelihood methods for the case of large numbers of changepoints and/or small data lengths, with mBIC recommended for smaller numbers of changepoints and/or longer lengths of data.


\bibliographystyle{apalike} 
\bibliography{elsbibfile}

\end{document}